\documentclass{an}
\usepackage{graphicx}
\usepackage{times}
\usepackage{fancyhdr}
\begin{document}

\title{Integrated Colours of Milky Way Globular Clusters and Horizontal 
Branch Morphology}

\author{G. H. Smith\inst{1} \and J. Strader\inst{1}}

\institute{University of California Observatories/Lick Observatory,
Department of Astronomy and Astrophysics, University of California, 
Santa Cruz, 95064 USA}

\date{Received; accepted; published online}

\abstract{Broadband colours are often used as metallicity proxies in the study 
of extragalactic globular clusters. A common concern is the effect of 
variations in horizontal branch (HB) morphology --- the second-parameter effect
--- on such colours. We have used $UBVI$, Washington, and DDO photometry
for a compilation of over 80 Milky Way globular clusters to address this 
question. Our method is to fit 
linear relations between colour and [Fe/H], and study the correlations between 
the residuals about these fits and two quantitative measures of HB morphology. 
While there is a significant HB effect seen in $U-B$, for the commonly used 
colours $B-V$, $V-I$, and $C-T_1$, the deviations from the baseline 
colour-[Fe/H] relations are less strongly related to HB morphology. There may 
be weak signatures in $B-V$ and $C-T_1$, but these are at the limit of 
observational uncertainties. The results may favour the use of $B-I$ in studies
of extragalactic globular clusters, especially when its high [Fe/H]-sensitivity
is considered.
}

\correspondence{graeme@ucolick.org}
\maketitle

\section{Introduction}

Integrated photometry has become a widely-used tool for the investigation of
globular cluster (GC) systems in galaxies, dating back to the work of 
van den Bergh (1967), who demonstrated that a reddening-independent index 
formed from the $UBV$ colours of Milky Way GCs could be correlated with 
metallicity. Despite several associated problems, it seems likely that 
integrated colours will continue to be used as metallicity proxies for 
extragalactic globular cluster systems that are beyond the reach of 
spectroscopic study. It also seems likely that the globular clusters of the 
Milky Way will continue to be used as calibrators for such studies. 
Calibrations have been published in a number of photometric systems (e.g., 
Aaronson et al.~1978; Frogel, Persson, \& Cohen 1980; Bica \& Pastoriza 1983; 
Brodie \& Huchra 1990; Geisler \& Forte 1990; Cohen \& Matthews 1994). For the 
purposes of studying globular clusters in M31, Barmby et al.~(2000) derived 
observational colour-metallicity relations for Milky Way GCs in a variety of 
broadband colours; these relations are in wide use.

There have also been a number of efforts to reproduce the integrated colours 
of Milky Way globular clusters with stellar population models, since these 
clusters form one of the best ``control'' samples for such tests. For example, 
Barmby \& Huchra (2000) compared the integrated colours of globular clusters
in the Milky Way and M31 with population synthesis models of Bruzual \& Charlot
(as reported in Leitherer et al.~1996), and Kurth, Fritz-von Alvensleben, \& 
Fricke (1999), as well as updated versions of the Worthey (1994) models. They 
concluded that an age difference of 4 Gyr or more was needed between metal-rich
and metal-poor clusters in order to provide the best model fits to their 
colours, with the metal-rich clusters being systematically younger. Covino et 
al.~(1994) used theoretical stellar spectral energy distributions to model the 
integrated colours of star clusters, and to explore their sensitivity to 
metallicity, horizontal branch (HB) morphology, red giant mass loss, and 
stellar mass function. They concluded that integrated 
colours are more sensitive to the first two of these factors. Maraston (1998) 
used evolutionary population synthesis models to calculate the changes in 
integrated $UBV$ colours caused by adding blue horizontal branch stars to a 
15 Gyr solar abundance model population. She found both $B-V$ and $U-B$ to be 
altered by $\sim 0.10$ to 0.15 mag by the addition of a blue horizontal branch 
(BHB) population. Further modelling was done by Brocato et al.~(2000), who 
explored the sensitivity of integrated colours to both age and statistical 
fluctuations in the number of bright giants. The Brocato et al.~(2000) models 
show that, particularly at low metallicity, the integrated $U-B$, $B-V$ and 
$V-I$ colours are sensitive to the value of the Reimers mass-loss parameter 
$\eta$ for red giants, since this affects the masses and temperatures of stars
on the horizontal branch. A detailed discussion is given by Yi (2003).

Other theoretical studies have demonstrated that in addition to metallicity and
age, the integrated spectra and hydrogen line strengths of globular clusters
should be sensitive to the temperature distribution of horizontal branch stars,
particularly those hotter than the RR Lyrae variables (Lee, Yoon, \& Lee 2000; 
Maraston \& Thomas 2000). On the observational side, 
de Freitas Pacheco \& Barbuy (1995) showed that the strength of the H$\beta$
line in the integrated spectra of Milky Way GCs was a ``bivariate function of
metallicity and a parameter describing the horizontal branch morphology.''
They found that BHB stars can make ``a substantial contribution'' to the
H$\beta$ line-strength. This influence of HB stars has been a concern in the
interpretation of integrated spectra of globular clusters, in part because
of the difficulty of disentangling their effects from those of age differences
(e.g., Peterson et al.~2003; Schiavon et al.~2004). Modeling performed by 
Puzia et al.~(2005) and Maraston (2005) suggests that BHB stars in metal-rich 
clusters can increase the value of the Lick H$\beta$ index by 0.4 \AA\ or 
greater, reducing derived ages by 5 Gyr or more. Colours based on filter 
bandpasses containing significant Balmer line absorption
might therefore be sensitive to horizontal branch morphology.

The main question addressed in this paper is whether a dependence upon
horizontal branch morphology can be discerned in the integrated-light colours 
of Milky Way globular clusters. Our approach is to investigate whether the 
scatter in various integrated colour-metallicity relations is correlated with 
parameters describing the horizontal branch structure. In addition, we check 
for correlations with several other cluster properties, such as central 
concentration and stellar mass function. Much of the focus of this paper is on 
colour systems that include filters with passbands
in the violet or ultraviolet regions of the spectrum, where blue horizontal
branch stars are most likely to have an influence on integrated cluster
magnitudes. The colours that we investigate are those of the Johnson $UBV$,
the Cousins $VI$, the DDO, and Washington systems. 

\section{{\it UBVI} Colours}

\subsection{The Colour and Metallicity Data Base}

We have collected both $UBVI$ and DDO colours for the combined set of Milky 
Way globular clusters observed by McClure \& van den Bergh (1968; MV68) and 
Bica \& Pastoriza (1983; BP83). The primary source consulted for dereddened 
$(U-B)_0$, $(B-V)_0$, $(V-I)_0$ colours and associated reddenings is Reed, 
Hesser, \& Shawl (1988; RHS88). They employed a database of observed colours 
having been compiled by Reed (1985), who homogenised data from 18 different 
literature sources. The majority of the data consist of aperture photometry 
obtained with 1P21 photomultipliers, with several sources (Hamuy 1984, Hanes 
\& Brodie 1985) having used S20 photomultipliers in order to obtain $I$-band 
data. The resulting colours are on the $UBV$ system 
defined by the measurements of van den Bergh (1967) and the $VRI$ system 
defined by the measurements of Hamuy (1984). As shown by Racine (1973), the 
reddening ratio $E(U-B)/E(B-V)$ of globular clusters varies with the spectral 
type of the integrated light, increasing in value towards later spectral type. 
The primary objectives of Reed et al.~(1988) were to determine the 
relationships between various integrated colours as a function of cluster 
spectral type, and to thereby obtain values of various interstellar-reddening 
ratios as a function of spectral type. Reddenings and intrinsic colours were 
derived for each cluster in their sample, and we have adopted their values for 
the intrinsic colours. 

Reed et al.~(1988) did not present any detailed discussion of the uncertainties
in the $UBVI$ colours. Reed (1985) gives values of a quantity denoted 
$\langle$RESIDUALS$\rangle$ that reflects the scatter among the colour 
measurements of each cluster from his various sources. It is equivalent to 
$\langle \vert C_i - \langle C_i \rangle \vert \rangle$, where the brackets
refer to weighted averages, and $C_i$ to the colour measurement for a given 
cluster from source {\it i} after transformation onto a uniform scale. Hamuy 
(1984) has compared his $U-B$, $B-V$, $V-R$, and $R-I$ measurements with the 
colours from Harris \& Racine (1979), who in turn list averages from different 
sources, and the observations of Hanes \& Brodie (1985). He finds that 
systematic differences between these sources can range from 0.01 to 0.03 mag, 
with scatters of $\pm 0.04$ mag for the $UBV$ colours and $\pm 0.05$ mag for 
$VRI$.

Where available, metallicities were compiled from the homogeneous data set of 
Rutledge et al. (1997a,b). They are based on the combined 
strengths of the $\lambda$8498, $\lambda$8542, and $\lambda$8662
Ca~${\scriptstyle{\rm II}}$ triplet lines of red giant branch (RGB) stars that 
are probable GC members. Rutledge et al.~(1997a,b) followed the techniques of 
Armandroff \& Da Costa (1991) for measuring the equivalent widths
of these lines, correcting them to a reference point on the RGB 
equal to the $V$ magnitude of the horizontal branch, and correlating them 
against various [Fe/H] measurements. The metallicities given in column 5 
of Table 2 of Rutledge et al.~(1997b) were adopted in this study, this column 
containing their ``most probable'' [Fe/H] value on the scale of Zinn \& West 
(1984; ZW84). Since the Rutledge et al.~(1997b) metallicities are based on 
spectroscopy of individual cluster red giants (typically $\sim$ 10-20 per 
cluster, Rutledge et al. 1997a), they should not be subject to 
horizontal branch effects in the way that metallicity estimates based on 
integrated cluster-light techniques could be. Where not available from this 
source, values of [Fe/H] for the clusters in our compilation were adopted from 
Table 6 of ZW84.

Reddenings for the clusters included in this study are available from both Reed
et al.~(1988) and the 1999 version (designated elsewhere in this paper as H99) 
of the catalogue of Milky Way GC properties described by Harris (1996). We have
tended to use the more recent reddenings from H99 for the purposes of dividing 
our cluster sample up into different reddening groups.

\subsection{{\it UBVI} Colour-Metallicity Relations}

Plots of $(B-V)_0$, $(U-B)_0$, and $(V-I)_0$ against [Fe/H] are shown in 
Figures 1 through 3. Symbols used in these figures denote both reddening as 
taken from H99 (filled symbols for $E(B-V) \leq 0.3$, open symbols for 
greater reddening), and whether a cluster has a core-collapse structure 
(triangle) or not (circle). Linear regression fits of each colour $C$ versus 
[Fe/H] were computed using $\chi^2$ minimisation. Fits were made
by minimising unweighted colours. Values of the coefficients for a variety 
of fits of the form $C = \alpha + \beta{\rm [Fe/H]}$ are given in Table 1, 
together with the variances $\sigma (\alpha)$ and $\sigma (\beta)$, and
a measure of the standard deviation in the GC colour residuals about each fit, 
\begin{equation}
\sigma_C 
 = \sqrt{\frac{\Sigma_{n=1}^N (C_n - \alpha - \beta{\rm [Fe/H]}_n)^2}{N-2}},
\end{equation}
where $N$ is the number of clusters fitted, and $C_n$ and ${\rm [Fe/H]}_n$ are 
the colour and metallicity of the {\it n}th cluster (Press et al.~1999). 
Each colour-metallicity fit is labelled with a letter 
in the first column of Table 1, and the eighth column specifies the clusters 
used in computing that fit. 

\begin{figure}
\resizebox{\hsize}{!}
{\includegraphics[]{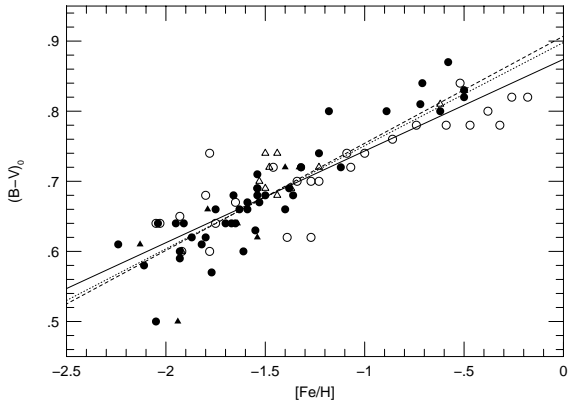}}
\caption{The integrated $(B-V)_0$ colour of Milky Way globular clusters plotted
versus metallicity [Fe/H]. Triangles represent core-collapse clusters, filled 
and open symbols correspond to clusters with reddenings of 
$E(B-V)_{\rm H99} \leq 0.3$ and $E(B-V)_{\rm H99} > 0.3$ respectively. Three 
different linear least-squares fits are shown: a fit made to all clusters in 
the sample (solid line); only clusters with $E(B-V)_{\rm H99} \leq 0.5$ (dotted
line); and only those clusters with both $E(B-V)_{\rm H99} \leq 0.5$ and [Fe/H]
$< -0.5$ (dashed line).
}
\label{Fig1}
\end{figure}

\begin{table*}
\caption{Coefficients in colour-metallicity relations: 
${\rm Colour} = \alpha + \beta {\rm [Fe/H]}$}
\label{tab1}
\begin{center}
\begin{tabular}{ccrcrccl}\hline
Fit               &
Colour            &
$\alpha$          &
$\sigma(\alpha)$  &
$\beta$           &
$\sigma(\beta)$   &
$\sigma_C$\footnotemark[1]     &
Sample \\
\hline
 A  & $(U-B)_0$  &  0.411  &  0.014  &  0.2032  &  0.0095  &  0.046  &   all clusters from MV68 and BP83 except NGC 6316  \\
 B  & $(U-B)_0$  &  0.434  &  0.017  &  0.2171  &  0.0111  &  0.044  &   $E(B-V)_{\rm H99} \leq 0.5$ (which also excludes NGC 6316) \\
 C  & $(U-B)_0$  &  0.441  &  0.018  &  0.2215  &  0.0115  &  0.044  &   $E(B-V)_{\rm H99} \leq 0.5$ and ${\rm [Fe/H]} \leq -0.5$   \\
 D  & $(B-V)_0$  &  0.874  &  0.011  &  0.1307  &  0.0078  &  0.039  &   all clusters from MV68 and BP83                  \\
 E  & $(B-V)_0$  &  0.898  &  0.014  &  0.1474  &  0.0094  &  0.037  &   $E(B-V)_{\rm H99} \leq 0.5$                                \\
 F  & $(B-V)_0$  &  0.907  &  0.015  &  0.1530  &  0.0095  &  0.036  &   $E(B-V)_{\rm H99} \leq 0.5$ and ${\rm [Fe/H]} \leq -0.5$   \\
 G  & $(V-I)_0$  &  1.097  &  0.023  &  0.1089  &  0.0157  &  0.076  &   all clusters from MV68 and BP83                  \\
 H  & $(V-I)_0$  &  1.124  &  0.023  &  0.1324  &  0.0148  &  0.056  &   $E(B-V)_{\rm H99} \leq 0.5$                                \\
 I  & $(V-I)_0$  &  1.164  &  0.025  &  0.1583  &  0.0162  &  0.049  &   $E(B-V)_{\rm H99} \leq 0.3$                                \\
 J  & $(B-I)_0$  &  2.373  &  0.038  &  0.4393  &  0.0261  &  0.126  &   all clusters from MV68 and BP83                  \\
 K  & $(B-I)_0$  &  2.501  &  0.037  &  0.5268  &  0.0245  &  0.092  &   $E(B-V)_{\rm H99} \leq 0.5$ and ${\rm [Fe/H]} \leq -0.5$   \\
{\it a}  & $C_0(42-45)$  &  0.808  &  0.010  &  0.1553  &  0.0066  &  0.033  &  excludes NGC 6144, 6325, 6517             \\
{\it b}  & $C_0(38-42)$  &  0.893  &  0.017  &  0.2291  &  0.0111  &  0.041  &  excludes NGC 6553                         \\
{\it c}  & $C_0(35-38)$  & --0.428 &  0.017  & --0.0803 &  0.0110  &  0.039  &  all clusters from MV68 and BP83           \\  
{\it d}  & $C_0(35-38)$  & --0.439 &  0.019  & --0.0897 &  0.0120  &  0.032  &  $E(B-V)_{\rm H99} \leq 0.3$                         \\
{\it e}  & $C_0(35-42)$  &  0.468  &  0.023  &  0.1508  &  0.0148  &  0.051  &  all clusters from MV68 and BP83           \\  
{\it f}  & $(C-T_1)_0$   &  1.727  &  0.029  &  0.3475  &  0.0194  &  0.076  &  all clusters from H\&C77                  \\
{\it g}  & $(C-M)_0$     &  0.984  &  0.019  &  0.2654  &  0.0127  &  0.050  &  all clusters from H\&C77                  \\
{\it h}  & $(M-T_1)_0$   &  0.743  &  0.011  &  0.0820  &  0.0075  &  0.029  &  all clusters from H\&C77                  \\
\hline
\end{tabular}
\end{center} 
$^1$ A measure of the standard deviation of cluster colour about the fitted colour-metallicity relation. The
subscript $C$ refers to the colour that is listed in the second column of each row.
\end{table*}

\begin{figure}
\resizebox{\hsize}{!}
{\includegraphics[]{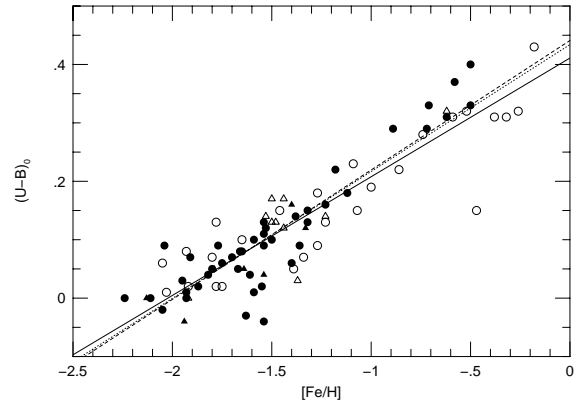}}
\caption{The integrated $(U-B)_0$ colour of Milky Way globular clusters plotted
versus metallicity [Fe/H]. Triangles represent core-collapse clusters, filled 
and open symbols correspond to clusters with reddenings of 
$E(B-V)_{\rm H99} \leq 0.3$ and $E(B-V)_{\rm H99} > 0.3$ respectively. Three 
different linear least-squares fits are shown: a fit made to all clusters in 
the sample with the exception of NGC 6316 (solid line); only clusters with 
$E(B-V)_{\rm H99} \leq 0.5$ (dotted line); and only those clusters with both 
$E(B-V)_{\rm H99} \leq 0.5$ and [Fe/H] $< -0.5$ (dashed line).
}
\label{Fig2}
\end{figure}

In plots of $(B-V)_0$ and $(U-B)_0$ versus [Fe/H] there seems to be some
separation between clusters with $E(B-V) \leq 0.3$ and $E(B-V) > 0.5$ at
metallicities of ${\rm [Fe/H]} > -0.5$. Consequently, for these 
colours, fits were computed for several different samples: (i) no constraint 
on either $E(B-V)$ or [Fe/H], (ii) only clusters with reddenings from H99 of 
$E(B-V)_{\rm H99} < 0.5$, and (iii) only clusters with reddenings of 
$E(B-V)_{\rm H99} < 0.5$ and metallicity ${\rm [Fe/H]} < -0.5$. Subsets (ii) 
and (iii) differ only in the object NGC 5927, which is excluded from sample 
(iii) on account of a high metallicity. The cluster NGC 6316 has been excluded 
from all fits involving $(U-B)_0$ due to a discordant position in Figure 2. 
Fits based on subset (iii) will be less subject to any non-linearity 
that may exist at high [Fe/H] in the colour-metallicity relations.

\begin{figure}
\resizebox{\hsize}{!}
{\includegraphics[]{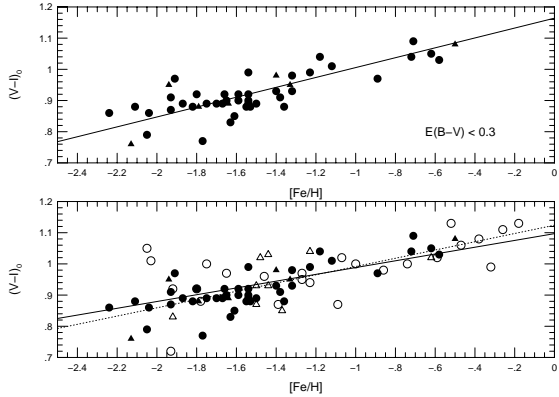}}
\caption{The integrated $(V-I)_0$ colour of Milky Way globular clusters versus
[Fe/H] metallicity. Symbols are the same as for Figures 1 and 2. The bottom 
panel shows the full cluster sample, while in the top panel only clusters with
reddenings $E(B-V)_{\rm H99} \leq 0.3$ are plotted. Linear least-squares 
regressions of $(V-I)_0$ versus [Fe/H] are shown for three different cluster 
sets: all clusters regardless of reddening (solid line in lower panel), 
clusters with reddening $E(B-V)_{\rm H99} \leq 0.5$ (dotted line in lower 
panel), and clusters with $E(B-V)_{\rm H99} \leq 0.3$ (solid line in top 
panel).
}
\label{Fig3}
\end{figure}

The bottom panel of Figure 3 shows all clusters with $(V-I)_0$ measurements,
while in the top panel only clusters with $E(B-V)_{\rm H99} \leq 0.3$ are 
shown. The scatter in $(V-I)_0$ at a given [Fe/H] is larger than for 
either $(U-B)_0$ or $(B-V)_0$. It is possible that this reflects  
larger observational uncertainties in $V-I$. Some evidence for this comes 
from the entries in the $\langle$RESIDUALS$\rangle$ column of Table III of 
Reed (1985). These residuals are stated to give ``an idea of the
uncertainties attendant to the final results.'' They are often 
higher for $V-I$ than for either $U-B$ or $B-V$, being 0.06-0.10 mag
for some clusters, consistent with $V-I$ being the most uncertain of the 
colours in Reed's homogenised data set.  However, in comparing his $V-R$ and 
$R-I$ photometry with that of Hanes \& Brodie (1985), Hamuy (1984) found a
scatter of $\pm 0.05$ mag, which is only slightly larger 
than the scatter between his $UBV$ colours and those compiled by Harris
\& Racine (1979). Three linear fits were made to the $(V-I)_0$ versus [Fe/H] 
relation: (i) all clusters regardless of reddening, (ii) clusters with 
reddenings of  $E(B-V)_{\rm H99} \leq 0.5$, 
and (iii) clusters with $E(B-V)_{\rm H99} \leq 0.3$. 

\begin{figure}
\resizebox{\hsize}{!}
{\includegraphics[]{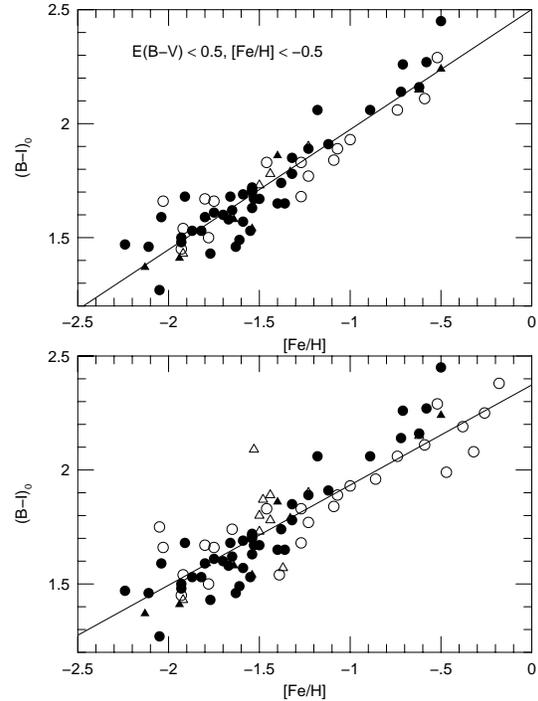}}
\caption{The integrated $(B-I)_0$ colour of Milky Way globular clusters versus
[Fe/H] metallicity. Symbols are the same as for Figures 1 and 2. The bottom 
panel shows the full cluster sample, while in the top panel only clusters with
reddenings $E(B-V)_{\rm H99} \leq 0.5$ and ${\rm [Fe/H]} \leq -0.5$ are 
plotted. Linear least-squares regressions of $(B-I)_0$ versus [Fe/H] are shown 
for each sample.
}
\label{Fig4}
\end{figure}
 
With regards to the $(U-B)_0$ and $(B-V)_0$ colours, the fits for the more 
restricted samples with lower reddenings have slightly greater 
slopes,\footnote{This may provide some evidence that the colour-[Fe/H] 
relations are slightly flatter at high metallicities, i.e., that they are 
non-linear. However, since many of the high-metallicity GCs are also highly 
reddened, it is not clear to what extent the differences in slope could be 
due to uncertainties in their intrinsic colours.} but the rms 
scatter about the fits is little different between the various cases.
In all three fits involving the $(V-I)_0$ colour the rms scatter is larger than
for either the $(U-B)_0$ or $(B-V)_0$ fits. The values of $\beta$ for the 
various fits indicate that $V-I$ and $B-V$ are similar in metallicity 
sensitivity. By comparison, the $(U-B)_0$ fits have both less scatter and 
greater metallicity sensitivity. If the greater scatter in the 
$(V-I)_0$-[Fe/H] relation is intrinsic to the clusters, rather than being due 
to observational errors, then the utility of this colour as a metallicity 
indicator would be somewhat compromised. 

The behaviour of $(B-I)_0$, calculated from the $(B-V)_0$ and $(V-I)_0$ values
of Reed et al.~(1988), is shown as a function of [Fe/H] in Figure 4. Symbols 
again denote reddening and core structure as in previous diagrams. In the lower
panel all clusters are plotted irrespective of reddening or metallicity; in the
upper panel only GCs with $E(B-V)_{\rm H99} \leq 0.5$ and [Fe/H] 
$\leq -0.5$ are presented. Linear least-squares fits are shown to both samples,
and the derived coefficients are listed in Table 1. 

\subsection{Residuals About the {\it UBVI} Colour-Metallicity Relations}

Colour residuals were measured with respect to the fits summarised in Table 1.
Throughout Section~2 we use the notation $\delta C_{\rm X}$ to refer to 
residuals that have been measured in a colour $C$ relative to fit X from
Table 1. Values of the Pearson linear correlation coefficient\footnote{This 
cofficient takes a value of 0.0 if there is no correlation between two 
variables, and values of $+1$ and $-1$ for perfect positive and negative 
correlations respectively.} $r$ (Press et al.~1999) between these colour 
residuals and a variety of GC parameters are listed in Table 2. These results 
are discussed in Sections 2.5 and 2.6. First, it is necessary to investigate 
whether the residuals may be affected by imperfectly known cluster reddenings.

The values of $\delta (V-I)_{\rm G}$ and $\delta (U-B)_{\rm A}$ are plotted 
versus $\delta (B-V)_{\rm D}$ in Figure 5. The residuals shown are those 
measured with respect to the least-squares fits labelled A, D and G in Table 1,
i.e., those fits made to the full sample of clusters unrestricted with respect 
to reddening or metallicity. The solid lines shown in Figure~5 pass through 
the origin and have slopes corresponding to reddening vectors of 
$E(U-B)/E(B-V) = 0.69$ and 0.96 (lower panel), and $E(V-I)/E(B-V) = 1.14$ and 
1.35 (upper panel). These reddening ratios correspond to the limits found by 
Reed et al.~(1988) for the earliest and latest spectral-type clusters. If the 
observed scatter about the various colour-metallicity fits were due entirely to
reddening errors, then clusters should fall between the solid lines in 
Figure~5. Given the addition of measurement errors of $\approx$ 0.04-0.05 mag
in the colours, there do seem to be trends in Figure~5 that could be partly 
accounted for by errors in cluster reddenings. We performed Spearman rank 
correlation tests (Spearman 1904) among the colour residuals. In the case of 
$\delta (U-B)_{\rm A}$ versus $\delta (B-V)_{\rm D}$, the resulting two-tailed 
$p$-value is $2 \times 10^{-9}$, while for $\delta (V-I)_{\rm G}$ versus 
$\delta (B-V)_{\rm D}$ we get $p = 2 \times 10^{-5}$. In general, $p$-values 
less than about 0.01 to 0.05 are considered to be significant. The value of the
Pearson $r$ parameter is 0.556 and $0.184$ respectively. Thus, there does 
appear to be a correlation between $\delta (U-B)_{\rm A}$ and 
$\delta (B-V)_{\rm D}$ that could be consistent, at least for some clusters, 
with errors in reddening. The situation for $\delta (V-I)_{\rm G}$ versus 
$\delta (B-V)_{\rm D}$ is more ambiguous.

\begin{figure}
\resizebox{\hsize}{!}
{\includegraphics[]{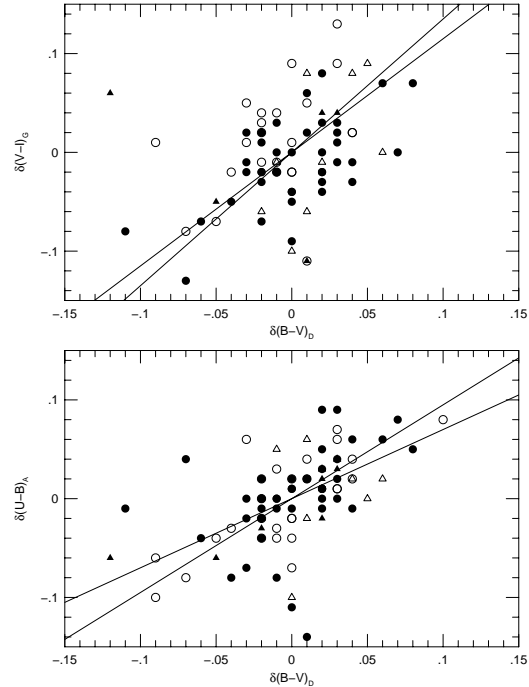}}
\caption{Relationships between residuals about three different 
colour-metallicity relations for Milky Way globular clusters. The linear 
least-squares fits about which these residuals were determined are based on 
the full sample of clusters from MV68 and BP83 unrestricted in metallicity or 
reddening. The coefficients of these fits are given in the rows labelled 
A, D, and G in Table 1. The solid lines in each panel have slopes equal to the
minimum and maximum values of the reddening ratios $E(U-B)/E(B-V)$ and 
$E(V-I)/E(B-V)$ found by Reed et al. (1988; RHS88) for Milky Way globular 
clusters. Symbols are the same as for Figure 1 and 2.
}
\label{Fig5}
\end{figure}

\begin{figure}
\resizebox{\hsize}{!}
{\includegraphics[]{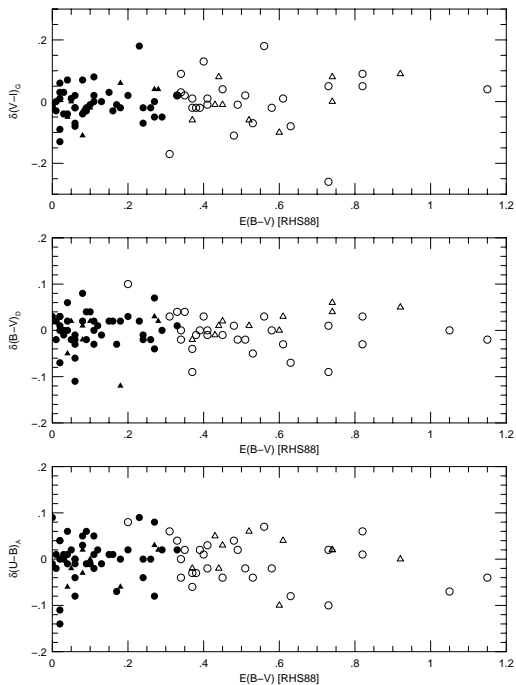}}
\caption{The colour residuals from Figure 5 are shown versus cluster
reddening from Reed et al.~(1988; RHS88). The residuals are with respect to 
the colour-metallicity fits listed as A, D, and G in Table 1. Symbols are the
same as for Figures 1 and 2.
}
\label{Fig6}
\end{figure}

\begin{table*}
\caption{Pearson correlation coefficients $r$}
\label{tab2}
\begin{center}
\begin{tabular}{llr}\hline
Correlation        &
Sample             &                      
$r$                \\
\hline
$\delta (U-B)_{\rm A}$ vs $B_T$            &  fit A; no [Fe/H] or $E(B-V)$ restrictions; 42 clusters                      &  --0.432   \\
$\delta (U-B)_{\rm C}$ vs $B_T$            &  ${\rm [Fe/H]} < -0.5$ and $E(B-V) < 0.5$; fit C; 41 clusters                &  --0.456   \\
$\delta (U-B)_{\rm A}$ vs $(B-V)_p$        &  fit A; no [Fe/H] or $E(B-V)$ restrictions; 42 clusters                      &  0.156     \\
$\delta (U-B)_{\rm C}$ vs $(B-V)_p$        &  ${\rm [Fe/H]} \leq -0.5$ and $E(B-V) \leq 0.5$; fit C; 41 clusters          &  0.015     \\
$\delta (U-B)_{\rm C}$ vs $\log \rho_0$    &  ${\rm [Fe/H]} \leq -0.5$ and $E(B-V) \leq 0.5$; fit C; 72 clusters          &  --0.110   \\
$\delta (U-B)_{\rm C}$ vs $c$              &  ${\rm [Fe/H]} \leq -0.5$, $E(B-V) \leq 0.5$, $c < 2.0$; fit C; 57 clusters  &  --0.019   \\
$\delta (U-B)_{\rm C}$ vs $b/a$            &  ${\rm [Fe/H]} \leq -0.5$ and $E(B-V) \leq 0.5$; fit C; 71 clusters          &  --0.197   \\
   
$\delta (B-V)_{\rm D}$ vs $B_T$            &  fit D; no [Fe/H] or $E(B-V)$ restrictions; 42 clusters                      &  0.006     \\
$\delta (B-V)_{\rm F}$ vs $B_T$            &  ${\rm [Fe/H]} \leq -0.5$ and $E(B-V) \leq 0.5$; fit F; 41 clusters          &  --0.004   \\
$\delta (B-V)_{\rm D}$ vs $(B-V)_p$        &  fit D; no [Fe/H] or $E(B-V)$ restrictions; 42 clusters                      &  0.406     \\
$\delta (B-V)_{\rm F}$ vs $(B-V)_p$        &  ${\rm [Fe/H]} \leq -0.5$ and $E(B-V) \leq 0.5$; fit F; 41 clusters          &  0.267     \\
$\delta (B-V)_{\rm F}$ vs $\log \rho_0$    &  ${\rm [Fe/H]} \leq -0.5$ and $E(B-V) \leq 0.5$; fit F; 72 clusters          &  --0.264   \\
$\delta (B-V)_{\rm F}$ vs $c$              &  ${\rm [Fe/H]} \leq -0.5$, $E(B-V) \leq 0.5$, $c < 2.0$; fit F; 57 clusters  &  --0.089   \\
$\delta (B-V)_{\rm F}$ vs $b/a$            &  ${\rm [Fe/H]} \leq -0.5$ and $E(B-V) \leq 0.5$; fit F; 71 clusters          &  --0.230   \\   

$\delta (V-I)_{\rm G}$ vs $B_T$            &  fit G; no [Fe/H] or $E(B-V)$ restrictions; 42 clusters                      &  --0.182   \\ 
$\delta (V-I)_{\rm I}$ vs $B_T$            &  $E(B-V) \leq 0.3$; fit I; 34 clusters                                       &  --0.293   \\ 
$\delta (V-I)_{\rm G}$ vs $(B-V)_p$        &  fit G; no [Fe/H] or $E(B-V)$ restrictions; 42 clusters                      &  0.185     \\
$\delta (V-I)_{\rm I}$ vs $(B-V)_p$        &  $E(B-V) \leq 0.3$; fit I; 34 clusters                                       &  0.016     \\
$\delta (V-I)_{\rm I}$ vs $\log \rho_0$    &  $E(B-V) \leq 0.3$; fit I; 49 clusters                                       &  --0.164   \\
$\delta (V-I)_{\rm I}$ vs $c$              &  $E(B-V) \leq 0.3$; fit I; 49 clusters                                       &  --0.156   \\
$\delta (V-I)_{\rm I}$ vs $c$              &  $E(B-V) \leq 0.3$ and $c < 2.0$; fit I; 38 clusters                         &  --0.323   \\ 
$\delta (V-I)_{\rm I}$ vs $b/a$            &  $E(B-V) \leq 0.3$; fit I; 49 clusters                                       &  0.079     \\

$\delta (B-I)_{\rm K}$ vs $B_T$         &  ${\rm [Fe/H]} \leq -0.5$ and $E(B-V) \leq 0.5$; fit K; 41 clusters              &  --0.325   \\
$\delta (B-I)_{\rm K}$ vs $B_T$         &  ${\rm [Fe/H]} \leq -0.5$, $E(B-V) \leq 0.5$, $B_T < 9.5$; fit K; 36 clusters    &  --0.286   \\  

$\delta C_0(42-45)$ vs $B_T$         &  fit {\it a}; no [Fe/H] or $E(B-V)$ restrictions; 43 clusters    &  --0.013   \\
$\delta C_0(42-45)$ vs $(B-V)_p$     &  fit {\it a}; no [Fe/H] or $E(B-V)$ restrictions; 43 clusters    &  0.210     \\
$\delta C_0(42-45)$ vs $(B-V)_p$     &  fit {\it a}; 40 clusters (excludes NGC 288, NGC 4147, NGC 5694) &  0.117     \\
$\delta C_0(42-45)$ vs $\log \rho_0$ &  fit {\it a}; no [Fe/H] or $E(B-V)$ restrictions; 88 clusters    &  0.027     \\
$\delta C_0(42-45)$ vs $c$           &  fit {\it a}; no [Fe/H] or $E(B-V)$ restrictions; 88 clusters    &  --0.142   \\
$\delta C_0(42-45)$ vs $b/a$         &  fit {\it a}; no [Fe/H] or $E(B-V)$ restrictions; 85 clusters    &  --0.074   \\

$\delta C_0(38-42)$ vs $B_T$         &  fit {\it b}; no [Fe/H] or $E(B-V)$ restrictions; 34 clusters    &  --0.543   \\
$\delta C_0(38-42)$ vs $(B-V)_p$     &  fit {\it b}; no [Fe/H] or $E(B-V)$ restrictions; 34 clusters    &  --0.074   \\
$\delta C_0(38-42)$ vs $\log \rho_0$ &  fit {\it b}; no [Fe/H] or $E(B-V)$ restrictions; 58 clusters    &  0.016     \\
$\delta C_0(38-42)$ vs $c$           &  fit {\it b}; no [Fe/H] or $E(B-V)$ restrictions; 58 clusters    &  0.007     \\
$\delta C_0(38-42)$ vs $b/a$         &  fit {\it b}; no [Fe/H] or $E(B-V)$ restrictions; 58 clusters    &  0.146     \\

$\delta C_0(35-38)$ vs $B_T$         &  fit {\it c}; no [Fe/H] or $E(B-V)$ restrictions; 32 clusters    &  0.163     \\
$\delta C_0(35-38)$ vs $B_T$         &  $E(B-V) \leq 0.3$; fit {\it d}; 26 clusters                     &  0.101     \\

$\delta C_0(35-42)$ vs $B_T$         &  fit {\it e}; no [Fe/H] or $E(B-V)$ restrictions; 32 clusters    &  --0.292   \\

$\delta (C-T_1)_0$ vs $B_T$          &  full sample (31 clusters)                                       &  --0.099   \\
$\delta (C-T_1)_0$ vs $(B-V)_p$      &  full sample (31 clusters)                                       &  0.361     \\
$\delta (C-T_1)_0$ vs $(B-V)_p$      &  excluding NGC 288, NGC 6218, NGC 6402 (28 clusters)             &  0.203     \\
$\delta (C-T_1)_0$ vs $\log \rho_0$  &  51 clusters                                                     &  0.069     \\
$\delta (C-T_1)_0$ vs $c$            &  51 clusters                                                     &  0.071     \\
$\delta (C-T_1)_0$ vs $b/a$          &  50 clusters                                                     &  0.027     \\

$\delta (C-M)_0$ vs $B_T$            &  full sample (31 clusters)                                       &  --0.180   \\
$\delta (C-M)_0$ vs $B_T$            &  excluding NGC 288, NGC 6218, NGC 6402 (28 clusters)             &  --0.266   \\
$\delta (C-M)_0$ vs $(B-V)_p$        &  full sample (31 clusters)                                       &   0.361    \\
$\delta (C-M)_0$ vs $(B-V)_p$        &  excluding NGC 288, NGC 6218, NGC 6402 (28 clusters)             &   0.196    \\

$\delta (M-T_1)_0$ vs $B_T$          &  full sample (31 clusters)                                       &  --0.031   \\
$\delta (M-T_1)_0$ vs $(B-V)_p$      &  full sample (31 clusters)                                       &  0.331     \\
$\delta (M-T_1)_0$ vs $(B-V)_p$      &  excluding NGC 288, NGC 6218, NGC 6402 (28 clusters)             &  0.189     \\
\hline                         

\end{tabular}
\end{center}
\end{table*}

\begin{table*}
\caption{Colour residuals for some example clusters}
\label{tab3}
\begin{center}
\begin{tabular}{lclrrrrrrr}\hline
Cluster                        &
[Fe/H]                         &
HB$^{1,2}$                     &
$\delta (U-B)_{\rm A}$         &
$\delta (U-B)_{\rm B}$         &
$\delta (B-V)_{\rm D}$         &
$\delta (B-V)_{\rm E}$         & 
$\delta (V-I)_{\rm G}$         &
$\delta (V-I)_{\rm I}$         &
$\delta (B-I)_{\rm K}$         \\
\hline   
NGC 1851  &  --1.23  &  B   &   0.00  & --0.01  &   0.03  &   0.02  &   0.03  &   0.02  &   0.04  \\
NGC 2808  &  --1.36  &  B,G & --0.04  & --0.05  & --0.02  & --0.02  & --0.07  & --0.07  & --0.13  \\ 
NGC 4590  &  --2.11  &  G   &   0.02  &   0.02  & --0.02  & --0.01  &   0.01  &   0.05  &   0.07  \\
NGC 5272  &  --1.66  &      &   0.01  &   0.01  &   0.02  &   0.03  &   0.00  &   0.02  &   0.05  \\
NGC 6121  &  --1.27  &  B   & --0.06  & --0.07  & --0.09  & --0.09  &   0.01  &   ....  & --0.15  \\ 
NGC 6205  &  --1.63  &  G   & --0.11  & --0.11  &   0.00  &   0.00  & --0.09  & --0.08  & --0.18  \\
NGC 6229  &  --1.54  &  B,G & --0.14  & --0.14  &   0.01  &   0.01  &   0.06  &   0.07  & --0.06  \\
NGC 6341  &  --2.24  &  G   &   0.04  &   0.05  &   0.03  &   0.04  &   0.01  &   0.05  &   0.15  \\
NGC 6362  &  --1.18  &  B,  &   0.05  &   0.04  &   0.08  &   0.08  &   0.07  &   0.06  &   0.18  \\
NGC 6388  &  --0.74  &  B,G &   0.02  &   0.01  &   0.00  & --0.01  & --0.02  &   ....  & --0.05  \\
NGC 6441  &  --0.59  &  B,G &   0.02  &   0.00  & --0.02  & --0.03  & --0.01  &   ....  & --0.08  \\ 
NGC 6638  &  --1.00  &  G   & --0.02  & --0.03  &   0.00  & --0.01  &   0.01  &   ....  & --0.04  \\
NGC 6681  &  --1.64  &  G   & --0.03  & --0.03  & --0.02  & --0.02  & --0.03  & --0.01  & --0.06  \\  
NGC 6712  &  --1.07  &  B   & --0.04  & --0.05  & --0.01  & --0.02  &   0.04  &   ....  & --0.05  \\
NGC 6723  &  --1.12  &  B   &   0.00  & --0.01  & --0.01  & --0.01  &   0.03  &   0.02  &   0.00  \\
NGC 6752  &  --1.54  &  G   & --0.06  & --0.06  & --0.05  & --0.05  & --0.05  & --0.04  & --0.15  \\
NGC 6809  &  --1.80  &  G   &   0.00  &   0.01  & --0.02  & --0.09  &   0.02  &   0.04  &   0.04  \\  
NGC 6864  &  --1.32  &  B   &   0.01  &   0.00  &   0.02  &   0.02  &   0.03  &   0.02  &   0.04  \\
NGC 7006  &  --1.59  &      & --0.08  & --0.08  & --0.01  &   0.00  & --0.02  & --0.01  & --0.09  \\ 
NGC 7078  &  --2.13  &  G   &   0.02  &   0.03  &   0.01  &   0.03  & --0.11  & --0.07  & --0.01  \\
\hline
\end{tabular}
\end{center} 
$^1$ B: denotes a cluster with a bimodal HB, i.e., a deficiency of RR Lyrae stars but well-populated
        blue and red horizontal branches.\\
$^2$ G: a cluster having a gap in the colour distribution of the blue horizontal branch stars. 
\end{table*}

The colour residuals from Figure 5 are plotted versus the reddening from
Reed et al.~(1988) in Figure 6, the symbols being the same as in previous 
figures. There seems to be no correlation between $E(B-V)_{\rm RHS88}$ and 
either the $\delta (B-V)_{\rm D}$ residual (a Spearman test gives a two-tailed 
$p$-value of 0.59; Pearson $r = -0.006$), or $\delta (U-B)_{\rm A}$ 
(Spearman $p = 0.98$, Pearson $r = -0.067$), with the possible exception of 
the two most-highly-reddened clusters. In the case of the 
$\delta (V-I)_{\rm G}$ residuals, the clusters with $E(B-V)_{\rm RHS88} > 0.6$ 
do seem to be systematically redder than would be expected from the 
least-squares fit (see Figure 6). The $p$-value from the Spearman rank 
correlation test of $\delta (V-I)_{\rm G}$ versus $E(B-V)_{\rm RHS88}$ is 
0.046, on the verge of significance, although the Pearson $r = 0.051$ does not
indicate a correlation. In the following discussions we often restrict 
consideration of $\delta (V-I)$ residuals to less-reddened systems.

In an effort to obtain some measure of the uncertainty in the reddenings 
of each cluster in the combined MV68 and BP83 samples, we calculated the 
differences $\Delta E(B-V)$ between the reddening values from RHS88 and H99. 
The values of $\Delta E(B-V)$ fall in the range --0.05 to 0.10 for all but 
three clusters in the MV68+BP83 sample. Spearman rank correlation tests 
between $\Delta E(B-V)$ and the colour residuals from Figure 6 were performed. 
These yielded $p$-values of 0.49, 0.08, and 0.195 for $\delta (U-B)_{\rm A}$,
$\delta (B-V)_{\rm D}$, and $\delta (V-I)_{\rm G}$ respectively. With the 
possible exception of a few clusters, there is no significant evidence from 
the Spearman test for any correlation between the $UBVI$ colour residuals and 
uncertainties in reddening. 

The data presented in Figures 5 and 6, together with the statistical tests 
noted above, therefore give somewhat ambivalent indications as to whether 
errors in measured cluster reddenings have a serious effect on the residuals 
about the $UBVI$ colour-metallicity relations of Figures 1-4. This complicating
effect will add scatter to any correlations we are searching for between these 
colour residuals and the horizontal branch morphology of globular clusters.  
In many of the discussions below we will accordingly conduct tests using both
the largest available sample of clusters, as well as samples restricted by
reddening. 

\subsection{$UBVI$ Colour Residuals for Some Example Clusters}

As noted above, one expected influence on integrated colours of globular 
clusters is the morphology of the horizontal branch. Values of $UBVI$ colour 
residuals are listed in Table~3 for some clusters with notable HB morphologies.
The pair of clusters M3 
(NGC 5272) and M13 (NGC 6205) constitute a classic example of the
second-parameter phenomenon (e.g., Catelan \& de Freitas Pacheco 1995).
Although they have very similar [Fe/H] abundances they differ in horizontal
branch morphology, with M13 having a more extensive blue horizontal branch
that extends to a $V$ magnitude comparable to the main sequence turnoff
(Johnson \& Bolte 1998; Rey et al. 2001). By contrast, NGC 7006 is similar
to both M3 and M13 in metallicity, but has a much larger fraction of HB stars
on the red side of the RR Lyrae gap (Sandage \& Wildey 1967; Buonanno et 
al.~1991). It is again a classic example of a second-parameter phenomenon.

In addition to the second-parameter triplet of M3-M13-NGC7006, $UBVI$ colour 
residuals are listed in Table 3 for other GCs having non-uniform horizontal 
branches. Even more extreme than the HB of M13 is that of NGC 2808, which not 
only extends to a $V$ magnitude below the main sequence turnoff, but has a 
deficiency of RR Lyrae variables and yet a well-populated red HB
(Harris 1974; Walker 1999; Bedin et al. 2000). As such, this cluster is said
to have a ``bimodal'' HB. In addition, there are other ``gaps'' in the colour 
distribution along the blue HB. NGC 6229 is another such bimodal HB 
cluster like NGC 2808; these clusters constitute two of the
cases in which gaps occur in the HB colour distribution both in the 
RR Lyrae region as well as along the BHB (Catelan et al. 1998).
NGC 6752 is a cluster that is known for an extended blue horizontal branch 
having considerable structure (e.g., Newell \& Sadler 1978, Buonanno 
et al.~1986, Momany et al.~2002). Both NGC 6388 and NGC 6441 are examples of
GCs with relatively high metallicities yet a population of blue HB stars
(Rich et al.~1997; Moehler, Sweigart, \& Catelan 1999). They are classified 
as being both ``bimodal'' and ``gap'' HB clusters by Catelan et al.~(1998).
Other GCs classified by Catelan et al.~(1998) as having either bimodal or gap 
HB morphologies are also listed in Table~3; these being identified by either 
the letter B or G in column 3. 

Are there clusters in Table 3 having integrated-colour residuals that might be 
attributed to the effect of a very blue horizontal branch? Possibly. Relative 
to M3 the extended-BHB cluster M13 has markedly bluer $(U-B)_0$ and $(B-I)_0$ 
colours, although the $(B-V)_0$ colours of these two objects are 
very similar\footnote{In the case of M13 $(U-B)_0 = -0.03$, $(B-I)_0 = 1.46$, 
and $(B-V)_0 = 0.66$, with equivalent colours of 0.08, 1.68, and 0.68 for M3.}.
The bimodal and BHB-gap cluster NGC 6229 also has a $(U-B)_0$ 
colour with a notable blue excess of $\delta (U-B)_{\rm A,B} = -0.14$,
although like M13, the $\delta (B-V)$ residual is not significantly 
different from zero. Oddly enough, NGC 7006 has an integrated $(U-B)_0$ 
colour that is only slightly redder than that of M13 despite the considerable 
difference in their HB structures. This can be seen from the
similar values of $\delta (U-B)_{\rm A,B}$ for these clusters (Table~3).
Noted for an extended BHB is NGC 6752, which does exhibit consistent blue 
excesses in all of the $UBVI$ colours (Table~3), as may NGC 2808. However, 
many of the clusters in Table 3 have colour residuals that do not depart 
significantly from zero, despite some notable properties to their horizontal 
branches. In the next section we look at possible connections between HB 
morphology and integrated colour in a more systematic way.

\subsection{Correlations Between {\it UBVI} Colour Residuals and Horizontal 
Branch Morphology}

Several parameters have been used in the literature to reflect the morphology 
of the horizontal branch. Two quantities, which we denote as $(B-V)_p$ and 
$B_T$, were introduced by Fusi Pecci et al.~(1993). The first of these 
parameters, $(B-V)_p$, refers to the dereddened colour at which the density of 
stars along the locus of the HB in a $V, (B-V)_0$ colour-magnitude diagram 
reaches a maximum. The second quantity is designed to reflect the extent 
of the blue horizontal branch. 
Formally, Fusi Pecci et al.~(1993) define $B_T$ as ``the length of 
the blue tail, measured along the ridge line of the HB starting from $(B-V)_p$
down to the adopted blue HB extreme.'' The parameter is measured along the
locus of the HB, and seems to be normalised in such a way that it has the 
value 10.0 for the horizontal branch of Messier 3. Examples of how $(B-V)_p$ 
and $B_T$ are defined can be found in Figure 3 of Fusi Pecci et al.~(1993). 
The $B_T$ parameter shows a wide dispersion 
at a given [Fe/H], particularly among intermediate-metallicity clusters. As 
discussed by Fusi Pecci et al.~(1993), this 
parameter can be used to study the systematics of second-parameter effects 
involving HB stars on the blue side of the RR Lyrae region. By contrast, 
$(B-V)_p$ is more nearly correlated with [Fe/H], although it still exhibits 
substantial scatter at a given metallicity indicative of second-parameter 
effects, particularly among clusters with [Fe/H] $\sim -1.5$.

\begin{figure}
\resizebox{\hsize}{!}
{\includegraphics[]{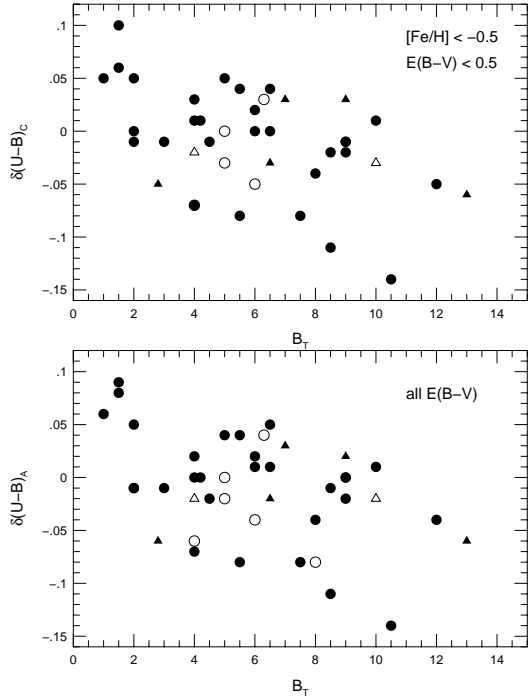}}
\caption{Two sets of $\delta (U-B)$ residuals versus the horizontal branch 
parameter $B_T$. The bottom panel shows residuals relative to the linear fit A 
from Table 1, which is based on all clusters in the present sample except 
NGC 6316. In the top panel the residuals are measured with respect to a linear
colour-metallicity relation based on the subset of clusters for which 
$E(B-V)_{\rm H99} \leq 0.5$ and ${\rm [Fe/H]} < -0.5$ (fit C of Table 1). 
Values of $B_T$ are not available for all of the clusters on which fits A and 
C are based. Symbols have the same meaning as in Figure 2.
}
\label{Fig7}
\end{figure}

\begin{figure}
\resizebox{\hsize}{!}
{\includegraphics[]{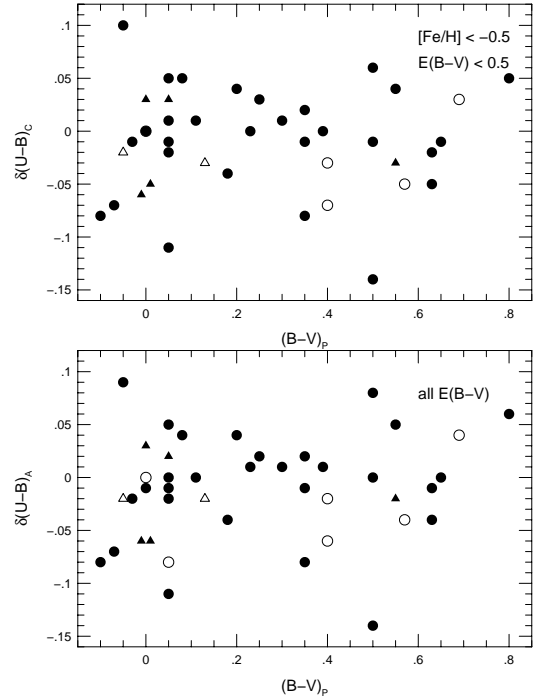}}
\caption{Two sets of $\delta (U-B)$ residuals versus the horizontal branch 
parameter $(B-V)_p$. Other details are the same as for Figure 7. Residuals
are relative to colour-metallicity fits A and C from Table 1. Symbols have the
same meaning as in Figure 2.
}
\label{Fig8}
\end{figure}

Two sets of $\delta (U-B)$ residuals are plotted versus HB parameters 
in Figures 7 and 8. Separate panels in each figure show residuals computed for 
the full sample of clusters (identified with fit A in Table 1), and for the 
subset restricted to $E(B-V)_{\rm H99} \leq 0.5$ and 
${\rm [Fe/H]} < -0.5$ (identified with fit C). There does appear to be a 
relation between $\delta (U-B)$ residuals and $B_T$, albeit with considerable 
scatter. If the unrestricted sample of clusters is considered, then the 
Pearson linear correlation coefficient between $B_T$ and the  
$\delta (U-B)_{\rm A}$ residuals is $r = -0.43$ (Table 2). There are 42 
clusters in this sample for which values of $B_T$ are available. The 
correlation coefficient is little different ($r = -0.46$ for 41 clusters) 
if the analysis is based on those GCs from the upper panel of 
Figure 7 together with fit C from Table 1. By contrast to these trends with 
$B_T$, Figure 8 shows little if any correlation between $\delta (U-B)$ and 
$(B-V)_p$, which is confirmed by the $r$-values listed in Table 2. 

In the case of the $B-V$ residuals there is little evidence for a relation with
the $B_T$ parameter, as the correlation coefficients in Table 2 indicate. This 
would suggest that $B-V$ is less susceptible to an extended blue HB morphology 
than $U-B$. By contrast, however, there may be a hint of some link between 
residuals in $(B-V)_0$ and the modal HB colour $(B-V)_p$, as illustrated in 
Figure 9. Adopting fit D from Table 1 and the full sample of clusters, the
Pearson correlation coefficient between $\delta (B-V)_{\rm D}$ and $(B-V)_p$ 
is $r = 0.41$ (Table 2). However, if instead we consider just those clusters
of restricted reddening and [Fe/H] for which fit F of Table 1 applies, then 
the resulting $\delta (B-V)_{\rm F}$ residuals and $(B-V)_p$ have a lower 
correlation coefficient of $r = 0.27$. Thus, the evidence for any relation 
between $(B-V)_0$ excess and HB morphology is sensitive to how the cluster 
sample is restricted with respect to reddening. Five of the six clusters in 
Figure 9 that have colour excesses of $\delta (B-V)_{\rm D,F} < -0.05$ do have
relatively blue modal HB colours of $(B-V)_p \leq 0.2$. Thus, there may be a
small subset of GCs whose $(B-V)_0$ colours show a second-parameter effect.
If $\delta(B-V)$ does have some correlation with $(B-V)_p$ but not $B_T$,
then it may imply either that $B-V$ is more sensitive to the mean HB colour
than to the colour distribution of extended blue-tail stars, or that a 
first-order fit to the $(B-V)_0$ versus [Fe/H] relation is not quite adequate.

\begin{figure}
\resizebox{\hsize}{!}
{\includegraphics[]{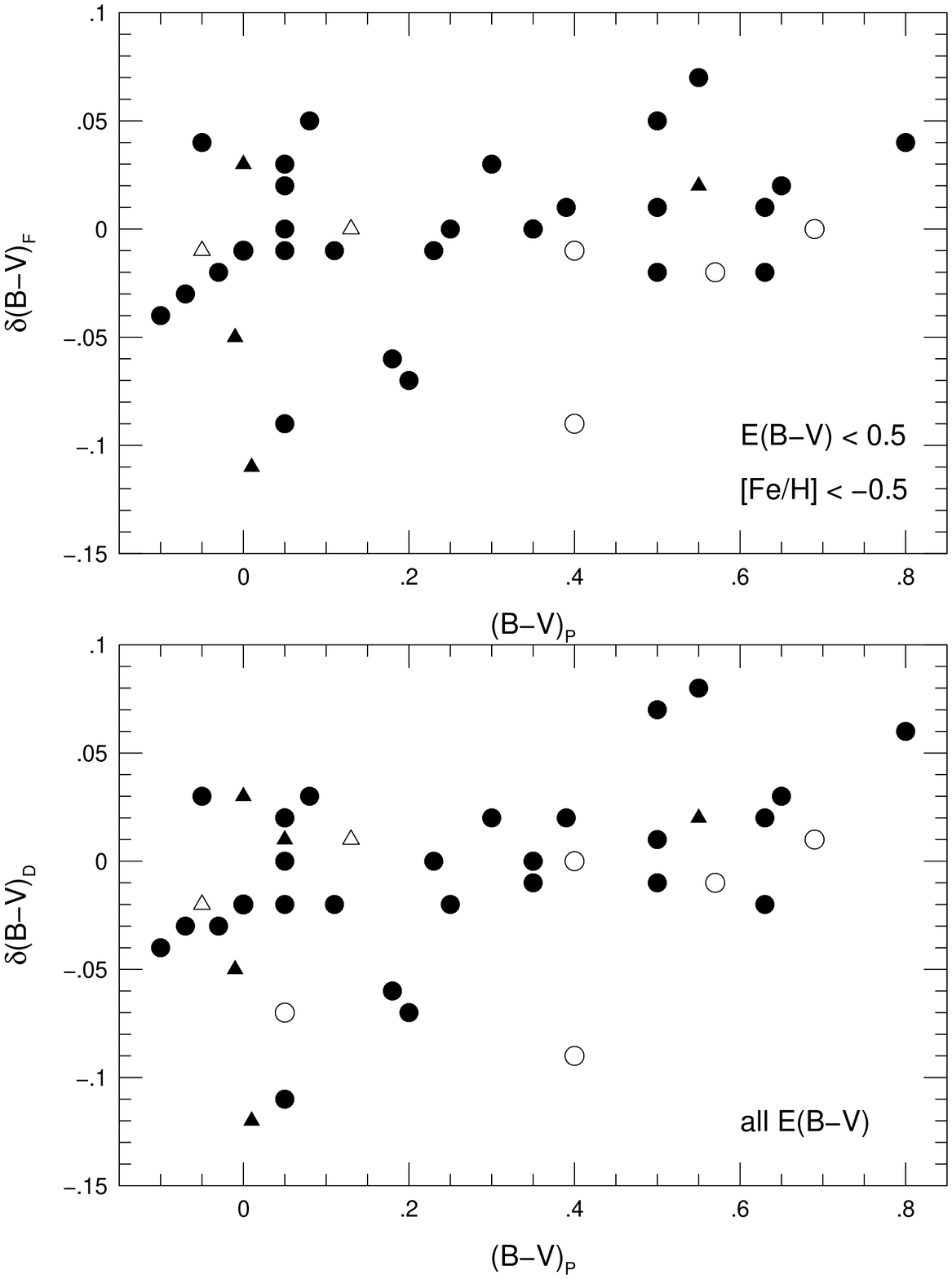}}
\caption {Two sets of $\delta (B-V)$ residuals versus the horizontal branch   
parameter $(B-V)_p$. The bottom panel shows residuals relative to the linear 
fit D from Table 1. In the top panel the residuals are measured with respect
to a linear colour-metallicity fit based on clusters for which
$E(B-V)_{\rm H99} \leq 0.5$ and ${\rm [Fe/H]} < -0.5$ (fit F of Table 1).
Symbols have the same meaning as in Figure 1.}
\label{Fig9}
\end{figure}

\begin{figure}
\resizebox{\hsize}{!}
{\includegraphics[]{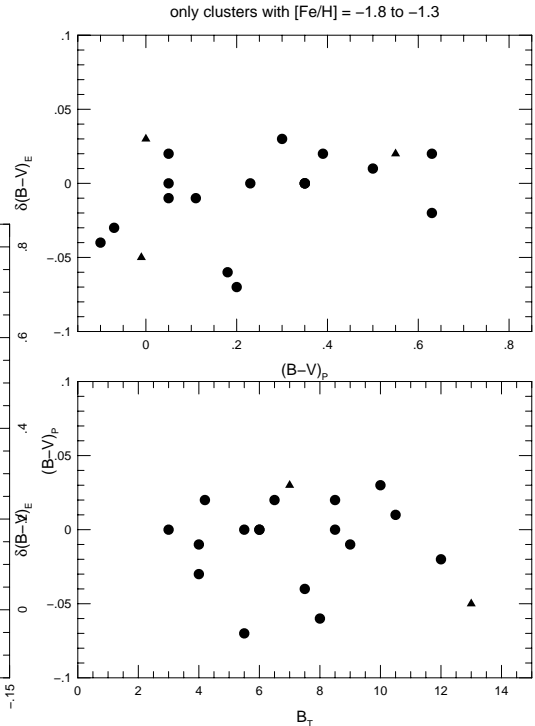}}
\caption{The $\delta (B-V)_{\rm E}$ residual versus both horizontal branch 
parameters $B_T$ and $(B-V)_p$. The residuals are measured with respect to
fit E from Table 1, and only clusters in the restricted metallicity
range $-1.8 \leq {\rm [Fe/H]} \leq -1.3$ are shown. Symbols have the same
meaning as in Figure 1.
}
\label{Fig10}
\end{figure}

Since the sensitivity of $(B-V)_p$ to second-parameter effects is most 
apparent among intermediate-metallicity clusters, Figure 10 shows the 
$\delta(B-V)_{\rm E}$ residuals (relative to fit E) versus both HB parameters 
for clusters in the metallicity range $-1.8 \leq {\rm [Fe/H]} \leq -1.3$. The 
residuals in this case are relative to the colour-metallicity relation based 
on clusters with $E(B-V)_{\rm H99} < 0.5$, but with no restriction on [Fe/H]. 
There is no correlation with $B_T$, as found for the full cluster sample. In 
the case of any correlation with $(B-V)_p$, Figures 9 and 10 show that this 
seems to be evinced by only a small number of clusters with relatively 
blue HB modal colours. All but four GCs in Figure 10 have 
$\vert \delta(B-V)_{\rm E} \vert \leq 0.03$, 
with the four most anomalous clusters having $(B-V)_p \leq 0.2$. Thus, 
among the intermediate-metallicity GCs, the second-parameter effect on 
integrated $(B-V)_0$ again seems to be limited to a small fraction of objects 
whose modal HB colour is fairly blue, although the values of $B_T$ for these 
clusters are not extreme (see the lower panel of Figure 10).

Sil'chenko (1983) used colour-magnitude diagrams of Milky Way globular
clusters from the Dudley catalogue (Philip, Cullen, \& White 1976) 
to compute the fraction of integrated cluster light in the 
$UBV$ passbands that is contributed by the horizontal branch population.  
One finding from this work was that for clusters of metallicity [Fe/H] $< -1.1$
the change in $B-V$ colour induced by subtracting out the horizontal branch 
stars is ``independent of the intrinsic horizontal-branch color,''  
although it does scale with the percentage contribution of HB stars to the
integrated $V$ magnitude. In other words, the effect of HB stars on the 
integrated $B-V$ colour was found to vary directly with the relative number of 
HB stars in a cluster, but was less sensitive to their colour distribution. 
Sil'chenko's (1983) findings may accord with the lack of any strong 
correlation between $\delta (B-V)$ and $B_T$, although the most discrepant 
clusters in Figures 9 and 10 may provide exceptions.

\begin{figure}
\resizebox{\hsize}{!}
{\includegraphics[]{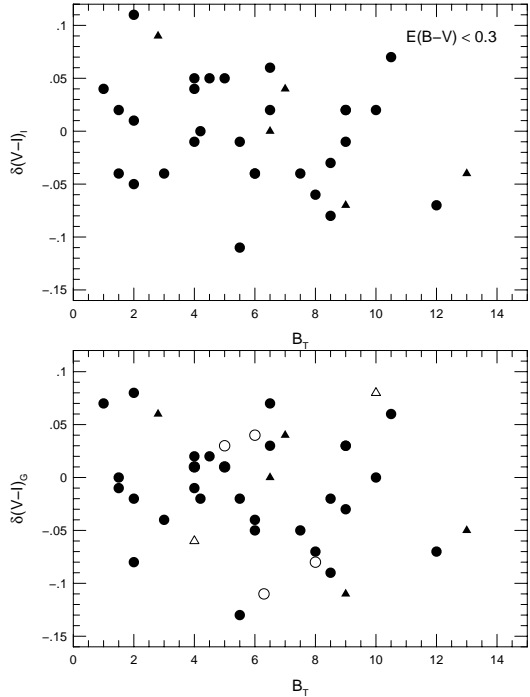}}
\caption{Two sets of $\delta (V-I)$ residuals versus the horizontal branch 
parameter $B_T$. In the top panel the residuals are measured with 
respect to a colour-metallicity relation based on the subset of 
clusters for which $E(B-V)_{\rm H99} \leq 0.3$ (fit I of Table 1). 
The bottom panel shows residuals relative to fit G, which is based on
all clusters with $V-I$ data regardless of reddening. Symbols have the
same meaning as in Figures 1 and 2.
}
\label{Fig11}
\end{figure}

Two sets of $\delta (V-I)$ residuals are plotted against $B_T$ in Figure 11.
There appears to be only weak evidence for a correlation. In the top panel 
there may be a very weak trend with considerable scatter between  
$\delta (V-I)_{\rm I}$ (relative to fit I) and $B_T$ for clusters with 
$E(B-V)_{\rm H99} < 0.3$ ($r = -0.29$; Table 2). However, this is not evident 
for the full sample of GCs unrestricted by reddening (bottom panel). There 
is no evidence of a correlation between the $\delta (V-I)$ residuals and 
$(B-V)_p$, as the correlation coefficients in Table 2 demonstrate. 

\begin{figure}
\resizebox{\hsize}{!}
{\includegraphics[]{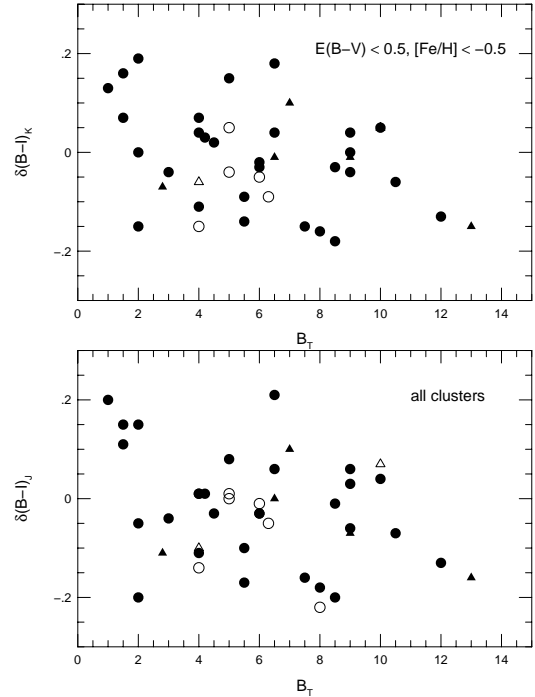}}
\caption{Two sets of $\delta (B-I)$ residuals versus the horizontal branch 
parameter $B_T$. The bottom panel shows results for the full sample of clusters
with $\delta (B-I)_{\rm J}$ residuals relative to the colour-metallicity 
relation given by fit J. Symbols are the same as for Figure 1. In the top panel
the residuals are measured with respect to fit K based on the subset of 
clusters for which $E(B-V)_{\rm H99} \leq 0.5$ and ${\rm [Fe/H]} \leq -0.5$. 
}
\label{Fig12}
\end{figure}

Could the lack of a correlation between $\delta (B-V)$ residuals and $B_T$ 
be due to comparable sensitivities of the integrated $B$ and $V$ magnitudes 
to extended blue horizontal branch morphology? If the $B$ magnitude is affected
by a long BHB tail then it may be more apparent in the $(B-I)_0$ colour, since 
the $I$ band samples less of the flux from BHB stars than the $V$ band. The 
$\delta (B-I)$ residuals about the colour-metallicity fits in Figure~4 are 
plotted versus the horizontal branch parameter $B_T$ in Figure 12. Among 
clusters with $B_T < 7$ there is basically a random scatter in
$\delta (B-I)$, while among those relatively few clusters with $B_T > 7$ 
there may be a trend to negative residuals, i.e., to bluer $(B-I)_0$ colours 
than given by the least-squares fits. The Pearson correlation cofficient 
between $B_T$ and $\delta (B-I)_{\rm K}$ for the clusters with [Fe/H] $< -0.5$ 
and $E(B-V)_{\rm H99} < 0.5$ is $r = -0.33$; if the sample is restricted to 
those clusters with $B_T < 9.5$ the coefficient is $r = -0.29$. This is 
comparable to the value of $r$ found for $\delta (V-I)_{\rm I}$ versus $B_T$. 
These results seem consistent with there being only a small influence of 
horizontal branch morphology on the $B-I$ and $V-I$ colours until an extended 
blue tail is developed. 

\subsection{Correlations with other Cluster Parameters?} 

\subsubsection{Cluster Structure}

Colour and stellar population gradients are found in some globular clusters. 
This is particularly the case with clusters having a high central concentration
or a post-core-collapse morphology, wherein the integrated cluster starlight 
becomes bluer towards the center in an effect that ``involves at least a few 
percent of the total visible light'' (Djorgovski et al.~1991). These colour 
gradients appear to be due to diminished numbers of red giants near cluster 
center (e.g., Howell, Guhathakurta, \& Tan 2000) possibly in combination with 
the presence of an inner population of stars that emits brightly at ultraviolet
wavelengths (Djorgovski et al.~1991; Djorgovski \& Piotto 1992). Mass 
segregation effects could therefore influence integrated colours measured 
within the inner regions of a globular cluster. This suggests checking whether 
the colour residuals calculated in the previous section show any trend with the
bulk physical properties of GCs.

We therefore tested for trends between the colour residuals 
$\delta (U-B)_{\rm A}$, $\delta (B-V)_{\rm D}$, and $\delta (V-I)_{\rm I}$
(measured with respect to fits A, D, and I from Table 1), versus several 
``structure'' parameters describing cluster mass, concentration, and
shape. The parameters chosen were: (i) the central density $\rho_0$ 
in units of $L_{\odot}$ pc$^{-3}$, (ii) the concentration parameter 
$c = \log r_t/r_c$, values for both of these being obtained from H99, and 
(iii) an ellipticity parameter $b/a$ measured by White \& Shawl (1987), which 
is equal to the ratio of the lengths of the minor and major axes as determined 
from cluster isophotes. Correlation coefficients calculated in these tests are 
listed in Table 2. In the case of the $UBV$ colour residuals there seem to be 
no discernible correlations with either $\log \rho_0$, $c$, or $b/a$. Any $UBV$
colour gradients that exist in core-collapse clusters do not seem to be 
detectable above the observational uncertainties in the integrated colours. 

\begin{figure}
\resizebox{\hsize}{!}
{\includegraphics[]{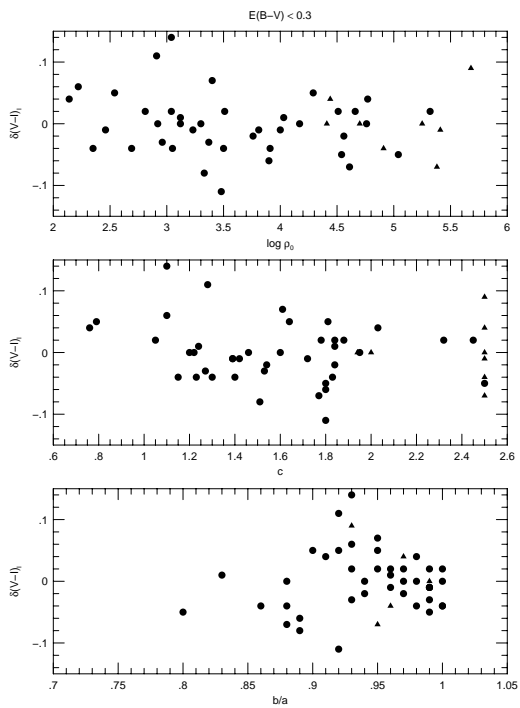}}
\caption{The colour residual $\delta (V-I)_{\rm I}$ measured relative to fit I 
from Table 1 versus GC central luminosity density 
$\log \rho_0$ ($L_{\odot}$ pc$^{-3}$), concentration parameter $c$, and 
ellipticity parameter $b/a$. Symbols are the same as for Figure 1.
}
\label{Fig13}
\end{figure}

The $\delta (V-I)_{\rm I}$ residuals are shown plotted against the three GC 
morphology parameters in Figure 13, the symbols again denote both reddening and
core structure as in previous plots. There may be some very modest trend with 
the concentration parameter $c$, in the sense that some clusters with $c < 1.3$
are slightly redder in $V-I$ than the mean colour-metallicity fit. However, 
among the core-collapse GCs there is a wide range in $\delta (V-I)_{\rm I}$,
with little propensity to the negative residuals that would be anticipated if 
the central regions of these clusters were depleted in red giants. For clusters
with $E(B-V)_{\rm H99} < 0.3$ and $c < 2.0$, the central concentration $c$ and 
the $\delta (V-I)_{\rm I}$ residuals have a Pearson correlation coefficient of
$r = -0.32$. Thus, the $V-I$ colour may show a marginal tendency to be redder 
in low-concentration clusters than in systems with $c > 1.3$. This could be 
consistent with some depopulation of the cores of high-concentration clusters 
in red giants. However, if such a phenomenon has occurred then it appears to 
be more detectable in $V-I$ than in $U-B$, and it is curious that the
integrated $V-I$ colours of core-collapse clusters with $c = 2.5$ do not
show any anomalies. In summary, with the possible exception of the $V-I$
residuals for low-reddening objects, there is little or no evidence for any
correlations between $UBVI$ colour residuals and globular cluster structure. 

\subsubsection{Cluster Age}

There is some evidence that for an age spread among Galactic globular 
clusters (e.g., Rosenberg et al.~1999). What effect might this have on our 
results? If age is strongly correlated with metallicity, then it should be 
accounted for in our initial colour-metallicity fits. However, if there is 
substantial scatter in cluster age at fixed metallicity, then age variations 
might be a factor in setting the colour residuals. To test for this, we have 
taken relative ages from De Angeli et al.~(2005) for 49 clusters in common 
with our sample. The relative age is defined as the ratio between the measured 
age and a fiducial average for GCs with metallicities on the Zinn \& West 
(1984) scale of [Fe/H] $< -1.7$. These clusters have a mean age of 11.2 Gyr. 
We performed Spearman rank correlation tests between the 
relative ages and both $\delta(U-B)_{\rm A}$ and $\delta(B-V)_{\rm D}$, 
two colour residuals that show at least partial evidence for dependence on the 
horizontal branch parameters $B_T$ and $(B-V)_p$ respectively. The resulting 
two-tailed $p$-values are 0.87 and 0.04 respectively. The Pearson $r$-values 
were found to be 0.025 and $-0.304$ respectively. These results indicate 
little or no evidence for a correlation in the case of the $U-B$ residual. We 
conclude that age variations are unlikely to be responsible for the correlation
found between the $U-B$ residuals and $B_T$. 

\begin{figure}
\resizebox{\hsize}{!}
{\includegraphics[]{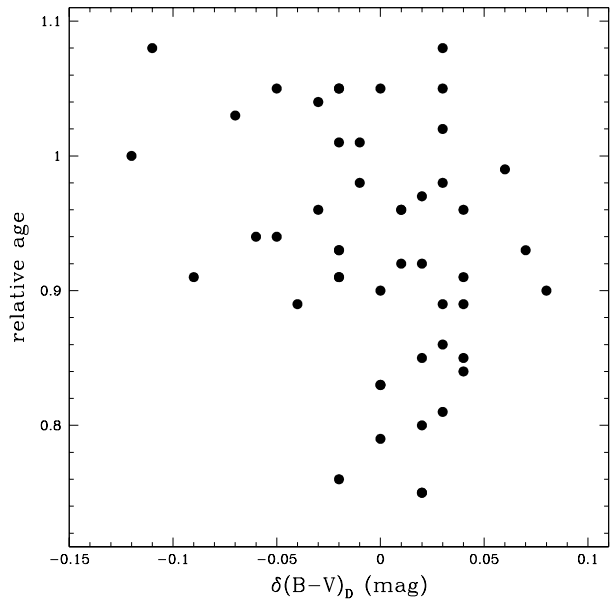}}
\caption{Relative cluster age, referenced to a mean of 11.2 Gyr
for metal-poor clusters with [Fe/H] $< -1.7$, versus the
$\delta (B-V)_{\rm D}$ colour residual. The ages are from De Angeli 
et al.~(2005).
}
\label{Fig14}
\end{figure}

A plot of relative age versus the $\delta(B-V)_{\rm D}$ residual is shown in 
Figure 14. The $p$-value from the Spearman test for this residual may suggest a
marginal correlation, as may the Pearson $r$-value. Inspection of Figure 14
indicates that this result may be weighted by a small subset of clusters with 
colour residuals of $\delta (B-V)_{\rm D} < -0.05$ and relative ages greater 
than 0.9. Such clusters also have fairly blue HB $(B-V)_p$ colours, as 
seen in Figures 9 and 10. Thus it may be difficult to disentangle an age from 
a second-parameter HB effect for these clusters. Any interpretation is further 
complicated by the fact that the relative ages of De Angeli et al.~(2005) are 
dependent on the magnitude difference between main sequence turnoff and 
horizontal branch stars, and so could be subject to second-parameter effects.

\section{DDO Colours}

Integrated photometry is also available for Milky Way globular clusters in the 
intermediate-band DDO filters. This system is well known to be useful for 
deriving photometric effective temperatures, gravities and 
metallicities of late-type stars, particularly red giants (e.g., Osborn 1973; 
Janes 1975; Sung \& Lee 1987; Claria et al. 1994a; Claria, Piatti, \& 
Lapasset 1994b). On account of the lower count levels with the DDO filters due 
to their narrower bandpasses and transmission in the violet region of the 
spectrum, integrated-light studies with this filter system have only been made 
to date for the Milky Way and M31 globular cluster systems.
 
Integrated DDO colours of Milky Way GCs have been measured by McClure \& 
van den Bergh (1968) and Bica \& Pastoriza (1983). The former observers
used one or two-channel photometers having 1P21 photomultiplier tubes on
the David Dunlap Observatory 1.9-m and Cerro Tololo Inter-American 
Observatory 0.4-m telescopes. Bica \& Pastoriza (1983) used a ``photon
counter device, equipped with an EMI9658 RAM photomultiplier'' on the 0.5-m
telescope of the Universidade Federal do Rio Grande do Sul. The DDO system 
incorporates five colours. The three more usual ones are denoted $C(45-48)$, 
$C(42-45)$ and $C(41-42)$. The first of these was only measured for Milky Way 
GCs by BP83, while both MV68 and BP83 give values of $C(42-45)$ and $C(41-42)$.
The $C(42-45)$ colour is the main effective temperature indicator in the DDO 
system, and was confirmed to be sensitive to cluster metallicity in integrated 
light by BP83 and Brodie \& Huchra (1990). In addition, MV68 measured 
two other colours: $C(38-41)$ and $C(35-38)$. The first of these is sensitive 
to the Ca~${\scriptstyle{\rm II}}$ H\&K break which is a known metallicity 
discriminator for galaxies. The second of these straddles the Balmer series 
break, with both the $C38$ and $C35$ filter bandpasses containing strong 
hydrogen lines. Brodie \& Huchra (1990) looked at the correlations between 
[Fe/H] and the integrated DDO colours $C(38-41)$, $C(38-42)$, $C(41-42)$, and 
$C(42-45)$ using the data of McClure \& van den Bergh (1968) for Milky Way 
globular clusters. They did not attempt to correct these colours for 
interstellar reddening.

McClure \& van den Bergh (1968) made multiple observations of many clusters, 
and their Table VI lists for each such object the average deviation of the 
individual measurements from the mean. These deviations vary considerably, 
being 0.02-0.03 in all colours for many clusters, but as large as 
0.06-0.10 for others. A comparison can be made between the measurements of 
MV68 and BP83 for the $C(42-45)$ colour. The mean value of the difference 
$\Delta C(42-45)$ [BP83--MV68] is 0.010 mag with a standard deviation of 
0.033 mag for the 64 clusters reported in both papers. We found no evidence
that this difference varies with cluster metallicity, with the Pearson
correlation coefficient between $\Delta C(42-45)$ and [Fe/H] being $r = 0.153$
for 64 clusters. 

Observed colours were corrected for reddening using the $E(B-V)$ values from
the H99 catalog, and adopting the standard reddening ratios of McClure \& 
van den Bergh (1968) and McClure (1973). Interstellar reddening in all of the 
DDO colours is small compared with $E(B-V)$. Given also the narrower DDO filter
bandpasses and smaller wavelength separation, the reddening ratios between DDO 
colours should show less of a variation with cluster spectral type 
than the $UBVI$ colours. We use the colour $C_0(38-42)$ rather than 
$C_0(38-41)$ throughout this paper. The $C41$ filter bandpass contains the 
$\lambda$4215 CN band, and since CN inhomogeneities are common within Milky Way
GCs we try to avoid this possible complication on the integrated colours. 
Nonetheless, the flux in the $C42$ filter is sensitive to the carbon abundance 
via the strength of the $\lambda$4300 G band (Bell, Dickens, \& Gustafsson 
1979; Tripicco \& Bell 1991). 

\begin{figure}
\resizebox{\hsize}{!}
{\includegraphics[]{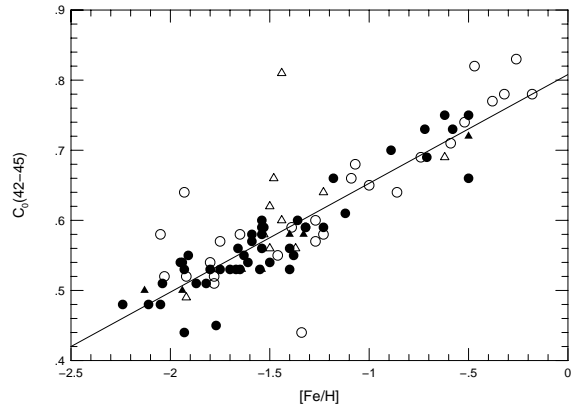}}
\caption{The integrated $C_0(42-45)$ colours of Milky Way globular clusters
versus metallicity. A linear least-squares fit (labelled {\it a} in Table 1) 
is shown. As with Figure 1, triangles denote core-collapse clusters, while 
filled and open symbols correspond to clusters with reddenings of 
$E(B-V)_{\rm H99} \leq 0.3$ and $E(B-V)_{\rm H99} > 0.3$ respectively. The 
three most discordant clusters in the figure were not included in the 
determination of the linear fit.
}
\label{Fig15}
\end{figure}

\begin{figure}
\resizebox{\hsize}{!}
{\includegraphics[]{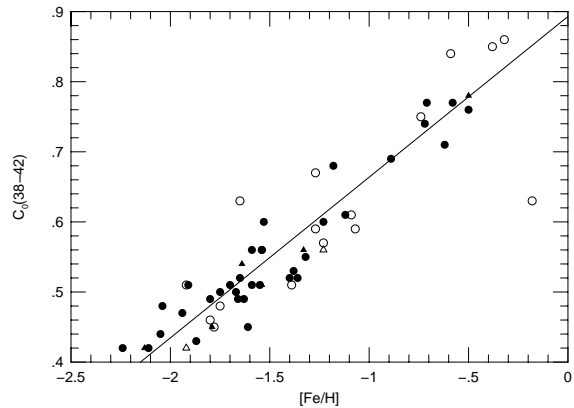}}
\caption{The integrated $C_0(38-42)$ colours of Milky Way globular clusters 
versus metallicity. A linear least-squares relation is shown, the coefficients 
for which are listed as fit {\it b} in Table 1. The discordant point for 
cluster NGC 6553 was not included in the fit. Symbols are the same as for 
Figures 1 and 15.
}
\label{Fig16}
\end{figure}

The most metallicity-sensitive of the standard DDO colours are $C_0(42-45)$ and
$C_0(38-42)$. Figures 15 and 16 show these colours versus [Fe/H]. In both cases
linear least-squares fits are shown as solid lines; they appear to match the 
observed correlations quite well. Coefficients of these fits are listed in 
Table 1 (fits {\it a} and {\it b}).
The cluster NGC 6553 was not included in the determination of the fit 
between $C_0(38-42)$ and [Fe/H], the point for this system being far removed 
from the mean trend in Figure 16. In the case of the $C_0(42-45)$ colour there 
are several clusters (NGC 6144, NGC 6325, and NGC 6517)
which fall well away from the main trend in Figure 15, and these have been
omitted in determining the fit coefficients listed in Table 1.
There are a number of clusters for which MV68 did not measure a $C(42-45)$
colour, such that the plotted values are based entirely on the 
measurements of BP83. NGC 6144, NGC 6325, and NGC 6517 are three such clusters,
and so without comparison observations, it is not clear whether their 
discordant positions in Figure 15 can be attributed to observational error.

\begin{figure}
\resizebox{\hsize}{!}
{\includegraphics[]{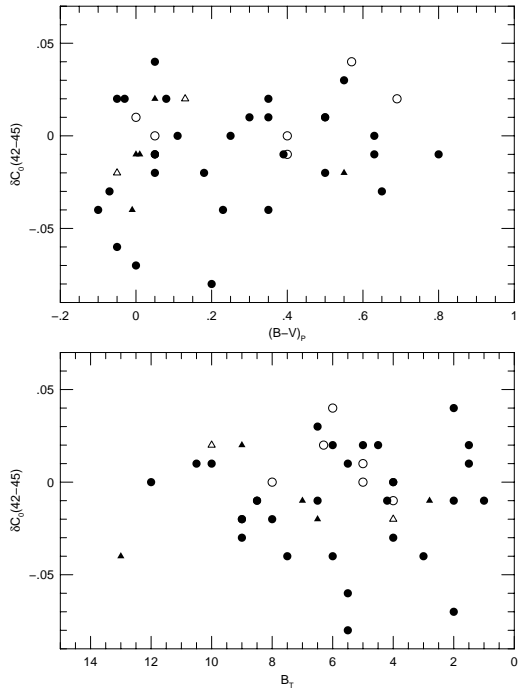}}
\caption{The $\delta C_0(42-45)$ residuals about the linear fit {\it a} shown
in Figure 15 versus the horizontal branch parameters $B_T$ and $(B-V)_p$.
Symbols have the same meaning as for Figure 1.
}
\label{Fig17}
\end{figure}

The $\delta C_0(42-45)$ residuals with respect to fit {\it a} are 
shown versus both horizontal branch parameters $B_T$ and $(B-V)_p$ in Figure 
17. There is no obvious correlation with the former of these indices, as 
confirmed by the Pearson $r$-value in Table 2. Also, with the possible 
exception of the three clusters with $\delta C_0(42-45) < -0.05$ (NGC 288, 
NGC 4147, and NGC 5694), there is no correlation of this colour excess with 
$(B-V)_p$. All three of these clusters have 
$(B-V)_p < 0.25$. However, many other globulars with similar HB modal colours 
are not abnormally blue in $C_0(42-45)$. In addition, these three clusters do 
not have extended BHB tails (they all have $B_T < 6.0$, 
see the lower panel of Figure 17). Given that all three objects lack 
photometry from McClure \& van den Bergh (1968), we feel that no compelling 
case can be made for a dependence of the $C_0(42-45)$ colour on horizontal 
branch morphology. The Pearson coefficients in Table 2 show that there are no 
correlations between the $\delta C_0(42-45)$ residuals and any of the cluster
structure parameters $\log \rho_0$, $c$, or $b/a$.

\begin{figure}
\resizebox{\hsize}{!}
{\includegraphics[]{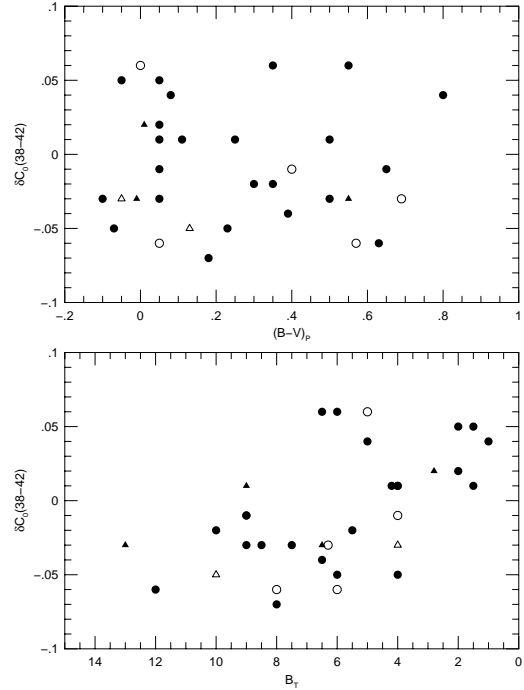}}
\caption{The $\delta C_0(38-42)$ residuals about the linear fit {\it b} from
Table 1 (see also Figure 16) versus the horizontal branch parameters $B_T$ and 
$(B-V)_p$. Symbols have the same meaning as for Figure 1.
}
\label{Fig18}
\end{figure}

Plots analogous to those of Figure 17, but for the residual 
$\delta C_0(38-42)$, are shown in Figure 18. The colour-metallicity 
relation for $C_0(38-42)$ reveals that this is a more sensitive [Fe/H] 
diagnostic than $C_0(42-45)$ for integrated light. The $\delta C_0(38-42)$ 
residual shows little trend with $(B-V)_p$ ($r = -0.07$), but it does
exhibit a correlation, albeit with considerable scatter, with the blue-tail 
parameter $B_T$ ($r = -0.54$, Table 2). There is no evidence for any relation 
between $\delta C_0(38-42)$ and either GC central density, concentration, 
or ellipticity (Table 2). 

As noted above, the $C_0(35-38)$ colour measures the strength of the Balmer 
jump. Consequently, it is sensitive to both stellar effective temperature and 
surface gravity. Within stars of the same luminosity class, this colour 
increases in value from spectral type B0 through A0, and then becomes more 
negative along a sequence from A through mid-K stars. Among A through M stars 
it increases in value as the luminosity class changes from V to III. 
The bandpass of the $C35$ filter is analogous to the Str\"{o}mgren $u$ filter 
(VB68), which is known to be sensitive to the strength of the $\lambda$3360 NH 
band in globular cluster red giants (Grundahl et al.~2002).

\begin{figure}
\resizebox{\hsize}{!}
{\includegraphics[]{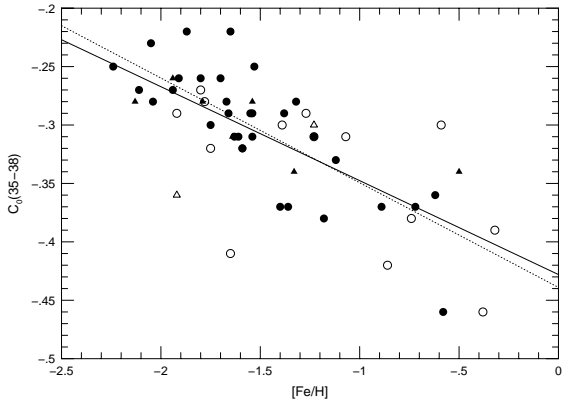}}
\caption{The integrated $C_0(35-38)$ colours of Milky Way globular clusters
versus metallicity. Symbols are the same as for Figure 1. Two linear 
least-squares fits are shown. The solid line gives the fit to all clusters 
(fit {\it c} in Table 1), whereas the dashed line shows the fit obtained when 
considering only clusters with reddenings of $E(B-V)_{\rm H99} \leq 0.3$ 
(fit {\it d}).
}
\label{Fig19}
\end{figure}

\begin{figure}
\resizebox{\hsize}{!}
{\includegraphics[]{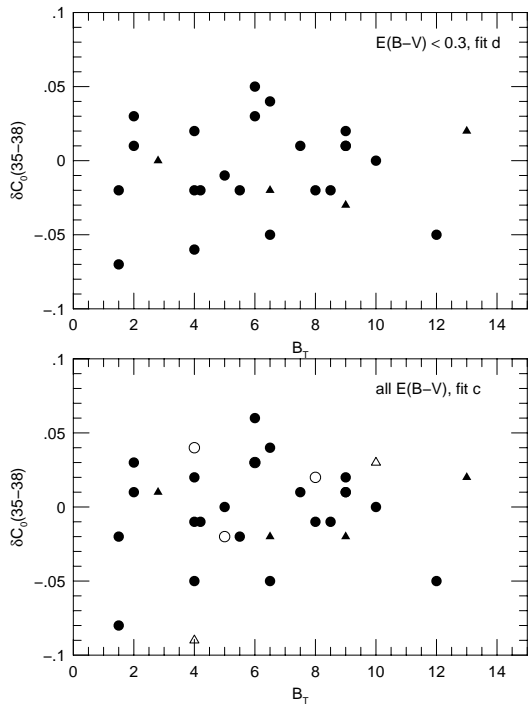}}
\caption{The $\delta C_0(35-38)$ residuals versus the horizontal branch
parameter $B_T$. Residuals about two different least-squares fits are shown,
one fit being derived from the full dataset of $C(35-38)$ colours 
(fit {\it c}), the other being based only on clusters with reddenings 
$E(B-V)_{\rm H99} \leq 0.3$ (fit {\it d}).
}
\label{Fig20}
\end{figure}

The behaviour of $C_0(35-38)$ versus [Fe/H] is shown in Figure 19. Two linear
least-squares fits were calculated, one using the entire cluster data set 
(fit {\it c} in Table 1), and a second restricted to clusters with reddening
$E(B-V)_{\rm H99} \leq 0.3$ (fit {\it d}). Both fits are shown in Figure 19. 
Although $C_0(35-38)$ shows a correlation with [Fe/H] it is not as sensitive as
the other DDO colours discussed above. Also in contrast to the previous 
colours, $C_0(35-38)$ becomes more negative with increasing [Fe/H], rather than
more positive. With increasing metallicity the giant branch and main sequence
turnoff becomes cooler, and the hydrogen lines from the cluster giants get 
weaker, thereby increasing the flux in the $C35$ filter relative to the $C38$ 
filter. As usual, $\delta C_0(35-38)$ residuals were calculated about each fit 
and are plotted versus the horizontal branch parameter $B_T$ in Figure 20. The 
notable result is that $\delta C_0(35-38)$ does not correlate with $B_T$ 
(see also Table 2). 

The lack of a dependence on HB morphology in the $C_0(35-38)$ colour appears to
be due to the $C35$ and $C38$ passbands both being sensitive to blue HB stars. 
Figure 21 shows the rather non-standard colour $C_0(35-42)$ versus [Fe/H], 
together with a linear least-squares fit (labelled {\it e} in Table 1) based on
all clusters with available data. There is a clear correlation with [Fe/H], but
it has a smaller slope than the $C_0(38-42)$ colour-metallicity relation. The 
$\delta C_0(35-42)$ residuals about fit {\it e} are plotted in Figure 22, where
they show some trend with $B_T$, although on account of some outliers the
resulting correlation coefficient of $r = -0.29$ is modest. The 
$\delta C_0(35-42)$ residuals at a given $B_T$ tend to be comparable to those 
in the $C_0(38-42)$ colour. Consequently, in forming the $C_0(35-38)$ colour, 
the HB effects in the $C35$ and $C38$ bands tend to cancel out, with 
little consequent dependence upon $B_T$ left. 

\begin{figure}
\resizebox{\hsize}{!}
{\includegraphics[]{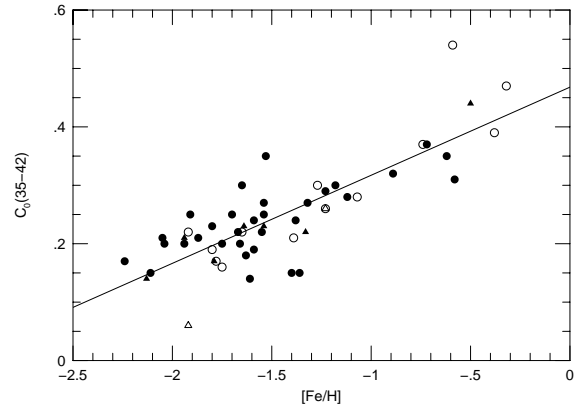}}
\caption{The integrated $C_0(35-42)$ colours of Milky Way globular 
clusters versus metallicity. Symbols are the same as for Figure 1. The 
linear least-squares fit (solid line) is based on all clusters for which 
data are available. The coefficients are listed as fit {\it e} in Table 1.
}
\label{Fig21}
\end{figure}

\begin{figure}
\resizebox{\hsize}{!}
{\includegraphics[]{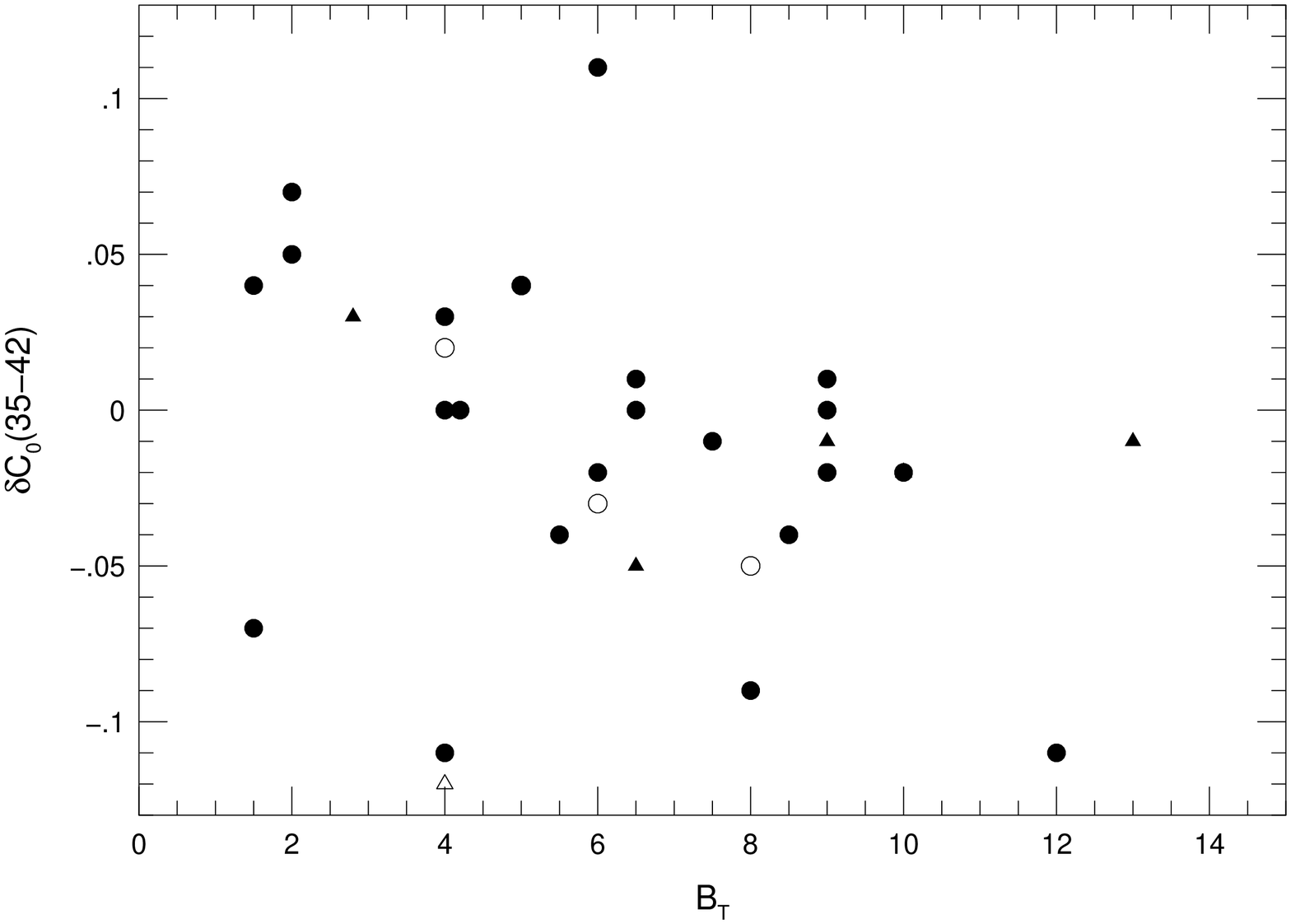}}
\caption{The $\delta C_0(35-42)$ residuals about fit {\it e} versus 
the horizontal branch parameter $B_T$. Symbols are the same as for Figure 1.
}
\label{Fig22}
\end{figure}

This cancellation effect is promoted by the presence of Balmer line absorption 
in both the $C35$ and $C38$ bandpasses. Also reducing the sensitivity of 
$C_0(35-38)$ to HB morphology could be the way in which Balmer line strengths 
vary with effective temperature along the horizontal branch. Figure 3 of 
McClure \& van den Bergh (1968) shows how $C_0(35-38)$ behaves as a function 
of spectral type between B0 and F0 for Population I dwarfs. This colour has a 
maximum value at A0, and is similar in value for F0 and mid-B spectral types. 
Thus, extending a horizontal branch hotter than an effective temperature of 
10,000 K would not cause $C_0(35-38)$ to become any redder, and may instead 
produce a degeneracy in colour as a function of $B_T$. 

\section{Washington Photometry}

The Washington system (Canterna 1976) has received substantial use in the study
of extragalactic globular clusters (e.g., Geisler \& Forte 1990; 
Harris et al. 1992; Lee \& Geisler 1993; McLaughlin et al.~1995; 
Ostrov, Forte, \& Geisler 1998; Forte et al.~2001; 
Harris, Harris, \& Geisler 2004). This system 
originally had two metallicity indicators denoted $M-T_1$ and $C-M$, the later 
of which is particularly useful for metal-poor stars (Geisler 1986; Geisler, 
Claria, \& Minniti 1991). Extragalactic globular cluster studies have tended 
to employ the $C-T_1$ colour as a metallicity index following the work of 
Geisler \& Forte (1990). Since the bandpass of the $C$ filter has an effective 
wavelength of 3910 \AA\ and a FWHM of 1100 \AA\ (Canterna 1976) it incorporates
a number of Balmer lines. In principle therefore both the integrated $C-M$ and 
$C-T_1$ colours could be sensitive to the HB morphology of globular clusters. 
To investigate this possibility, we have used the 
integrated $C-M$ and $M-T_1$ colours of 51 GCs from Harris \& Canterna (1977; 
H\&C77), combined with the H99 reddenings and the metallicities described 
above. The H\&C77 observations were obtained with telescopes at Manastash Ridge
Observatory, Kitt Peak National Observatory, and Cerro Tololo Inter-American
Observatory, equipped with pulse-counting photometers
using RCA Ga-As photomultiplier tubes.
Reddening ratios of $E(M-T_1)/E(B-V) = 0.95$ and $E(C-M)/E(B-V) = 1.14$ 
(Harris \& Canterna 1977) were used to obtain the intrinsic colours 
$(C-M)_0$, $(M-T_1)_0$, and $(C-T_1)_0$.

\begin{figure}
\resizebox{\hsize}{!}
{\includegraphics[]{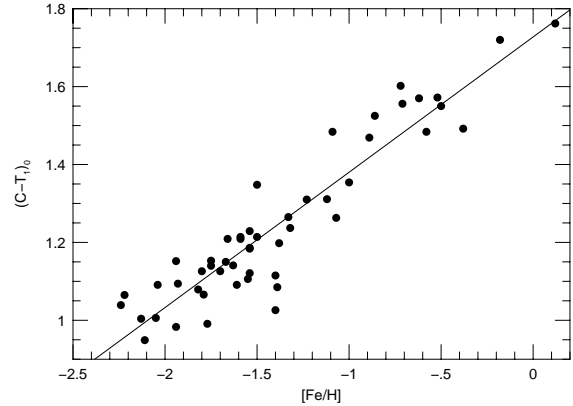}}
\caption{The integrated $(C-T_1)_0$ colours of Milky Way globular clusters
versus [Fe/H] metallicity. A linear least-squares fit ({\it f} in Table 1) is 
shown.
}
\label{Fig23}
\end{figure}

\begin{figure}
\resizebox{\hsize}{!}
{\includegraphics[]{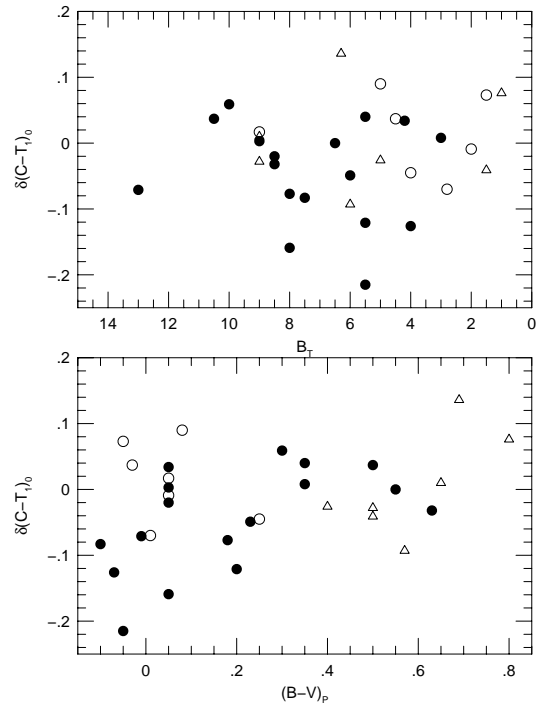}}
\caption{Residuals $\delta (C-T_1)_0$ relative to the least-squares fit {\it f}
versus the horizontal branch parameters $B_T$ and $(B-V)_p$. Symbols here 
differ from the conventions used in previous figures, and are based on 
metallicity: ${\rm [Fe/H]} < -1.8$ (open circles), 
$-1.8 \leq {\rm [Fe/H]} \leq -1.3$ (filled circles), 
${\rm [Fe/H]} > -1.3$ (open triangles).   
}
\label{Fig24}
\end{figure}

The $(C-T_1)_0$ versus metallicity diagram and a linear fit to it are shown 
in Figure 23. The coefficients of this fit, which was made to the full sample, 
are given in row {\it f} of Table 1. The colour residuals about the fit are 
plotted versus both HB parameters $B_T$ and $(B-V)_p$ in Figure 24. There may 
be a weak relation with $(B-V)_p$, but not with the blue-tail parameter $B_T$
(see Table 2). The value of the Pearson linear correlation coefficient between 
$\delta (C-T_{1})_{0}$ and $(B-V)_p$ is 0.36 for those 31 clusters that have 
both integrated Washington photometry and measured HB parameters. The symbols 
in Figure 24 refer to cluster metallicity, with filled circles denoting the 
metallicity range $-1.8 \leq {\rm [Fe/H]} \leq -1.3$ wherein $(B-V)_p$ shows 
the greatest scatter due to the second-parameter effect. The 
intermediate-metallicity clusters with HB modal colours of $(B-V)_p < 0.0$ do 
have relatively blue $(C-T_1)_0$ residuals, which may suggest that this 
Washington colour is sensitive to HB morphology if the blue side of the 
RR Lyrae gap is sufficiently populated. However, this tentative trend is 
heavily weighted by three clusters with [Fe/H] $\approx -1.4$, namely NGC 288, 
NGC 6218, and NGC 6402, for which $\delta (C-T_1)_0$ is equal to --0.215, 
--0.126, and --0.159 respectively, the lowest values of any clusters in the 
sample. They fall farthest below the linear-fit line in Figure 23. If they are 
excluded from the sample the Pearson correlation coefficient drops to 
$r = 0.20$. Correlation coefficients for the $\delta (C-T_{1})_{0}$ residuals 
versus the GC structure properties $\log \rho_0$, $c$, and $b/a$ are listed in 
Table 2. In all three cases $r < 0.1$, with no evident correlations.

The $(C-M)_0$ and $(M-T_1)_0$ colours are plotted versus [Fe/H] in Figure 25. 
Linear fits to the full data sets are shown, the coefficients of which are 
listed in Table 1. The corresponding $\delta (C-M)_0$ residuals are plotted 
versus $B_T$ and $(B-V)_p$ in Figure 26. The trends seen for $\delta (C-M)_0$ 
are much the same as for $\delta (C-T_1)_0$, there being no significant 
correlation with $B_T$ for the full sample of clusters, or with $(B-V)_p$ for 
clusters having horizontal branches with $(B-V)_p > 0$. There may be a tendency
for some intermediate-metallicity clusters with $(B-V)_p < 0.1$ to have a blue 
$(C-M)_0$ excess. However, as with the case with the $(C-T_1)_0$ colour, this 
tendency is due largely to the three clusters NGC 288, NGC 6218, and NGC 6402. 
If these three clusters are excluded, the correlation coefficient drops from 
$r=0.36$ to 0.20.

\begin{figure}
\resizebox{\hsize}{!}
{\includegraphics[]{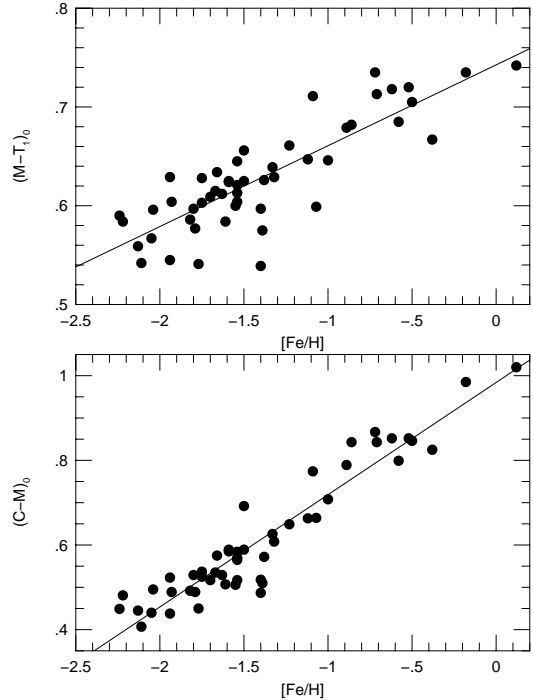}}
\caption{The integrated Washington $(C-M)_0$ and $(M-T_1)_0$ colours of Milky 
Way globular clusters are plotted versus metallicity. Linear least-squares fits
to each dataset are shown, with the coefficients listed as fits {\it g} and
{\it h} respectively in Table 1.
}
\label{Fig25}
\end{figure}

\begin{figure}
\resizebox{\hsize}{!}
{\includegraphics[]{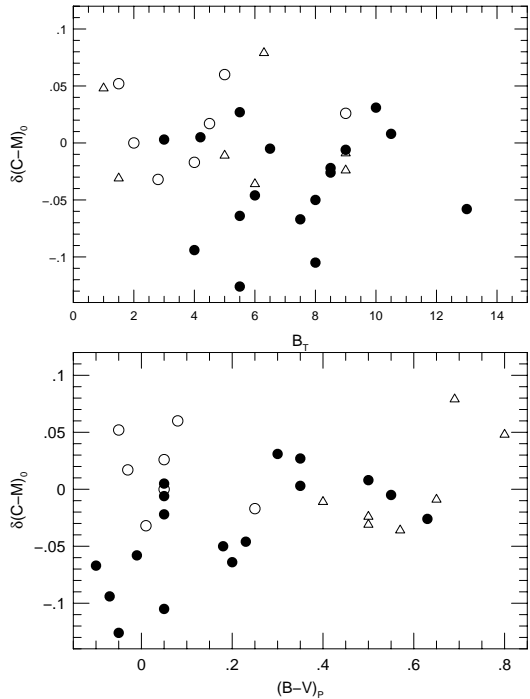}}
\caption{The $\delta (C-M)_0$ residuals about the linear fit {\it g} from 
Table 1 (also shown in Figure 25) versus the horizontal branch parameters 
$B_T$ and $(B-V)_p$. Symbols here are based on metallicity and have the same 
designation as for Figure 24: ${\rm [Fe/H]} < -1.8$ (open circles), 
$-1.8 \leq {\rm [Fe/H]} \leq -1.3$ (filled circles), 
${\rm [Fe/H]} > -1.3$ (open triangles).   
}
\label{Fig26}
\end{figure}

The sensitivity of $(M-T_1)_0$ to metallicity is much less than for either of 
the colours including the $C$ band, as can be seen from Figure 6 of Harris et 
al.~(1992). The $\delta (M-T_1)_0$ residuals behave similarly to those of the 
other Washington colours discussed above. There is no obvious correlation with 
$B_T$, and while there may be a slight trend with $(B-V)_p$, it is again 
strongly influenced by the three clusters NGC 288, NGC 6218, and NGC 6402 (see 
Table 2). In summary, the behaviour of the Washington colours could be similar 
to that of $B-V$, in that they may be more sensitive to the modal colour of 
the horizontal branch than to the extent of the BHB tail.

\section{Summary and Conclusions}

In summary, we find that the integrated $U-B$ colours of Milky Way globular 
clusters do exhibit a significant dependence on blue horizontal branch 
morphology as parameterised by the $B_T$ index. The situation for the $B-V$ 
colour is rather equivocal, with some very modest dependence possibly being 
discerned with respect to the HB parameter $(B-V)_p$, but not the blue-tail 
parameter $B_T$. However, this possible effect is rather sensitive to clusters 
of high reddening. Among intermediate-metallicity GCs, any significant 
second-parameter effect on integrated $(B-V)_0$ seems to be limited to a small 
subset of objects whose HBs have relatively blue modal colours but not 
necessarily extended blue tails. The $V-I$ colour evinces
considerable scatter about a mean colour-metallicity relation for the Milky
Way globulars, with the fit obtained being influenced by the extent to which
the most heavily reddened clusters are included in the sample. The origin of
the scatter in $V-I$ at a given [Fe/H] is not clear. There may be a modest
relation with the cluster concentration $c$ in the sense expected if some
high-concentration clusters are diminished in the number of red giants in 
their cores. However, any such intrinsic scatter is again at a level
comparable to the observational uncertainties, and furthermore there is no 
evidence of a difference in mean $UBVI$ colours between core-collapse 
and non-core-collapse clusters at a given metallicity. Evidence for any 
correlations between integrated colours and GC structure is weak at best.

Based on the values of $\sigma_C/\beta$ from Table 1, and the degree of
sensitivity (or otherwise) to horizontal branch morphology, we suggest that
the $B-V$ or $B-I$ colours may be preferable metallicity indicators to the 
often-employed Cousins $V-I$. The latter colour has the appeal that globular 
clusters have higher integrated fluxes in the $V$ and $I$ passbands than in 
the $B$ band. However, until the source of the relatively large scatter about 
the observed $(V-I)_0$-metallicity relation for Milky Way GCs can be 
pinned down, the application of this colour as a metallicity diagnostic 
warrants further investigation. 

On the basis of Figure 7, a globular cluster having a horizontal branch with a 
very elongated blue tail (say $B_T > 8$) could have a $(U-B)_0$ colour that is
$\Delta (U-B) = 0.05-0.10$ mag bluer than another GC with the same [Fe/H] 
but a less extreme BHB (with $B_T \sim 2$ to 4). If the $(U-B)_0$ colour were 
being used as a metallicity diagnostic, then the [Fe/H] of the former
cluster could be underestimated by $\Delta {\rm [Fe/H]} = \Delta (U-B)_0/\beta
= 0.25$ to 0.5 dex, based on the slope of fit A from Table 1. This rather
considerable effect illustrates the level to which the $(U-B)_0$ colour could 
be compromised for metallicity estimation by horizontal-branch effects.
In the case of the $(B-V)_0$ colour, there may be a small number of clusters
in Figures 9 and 10 which have $\delta (B-V)_0 < -0.05$ due to a blue 
horizontal branch. An effect of this magnitude would cause the $(B-V)_0$
colour to understimate the metallicity of a globular cluster by $> 0.4$
dex. Thus, although we find that any significant effect of HB morphology on
$(B-V)_0$ colour may be limited to a small fraction of GCs in the Milky Way,
if it does occur it can lead to quite substantial errors in the inferred
metallicity.

The DDO $C_0(38-42)$ colour is found to be sensitive to the blue tail of the
horizontal branch, as parameterised through the index $B_T$, although no such 
effect is seen in the $C_0(42-45)$ colour. Thus, as with the case for $U-B$, 
the HB blue tail seems to be most effective in modifying colours which have one
bandpass blueward of 4000 \AA. By contrast, in the case of the $C_0(35-38)$ 
colour, which has both bandpasses blueward of 4000 \AA, the effects of hydrogen
line absorption in each band partially cancel out, so that it exhibits 
less variation with $B_T$ than do either $C_0(38-42)$ or $C_0(35-42)$. In the 
case of the integrated $C_0(38-42)$ colour, extremes of BHB morphology might
cause an effect of $\sim 0.05$ mag, based on the lower panel of Figure 18.
This could lead to an error in [Fe/H] of $\sim 0.2$-0.25 dex if the colour 
was being used as a metallicity diagnostic. 

The Washington colours may be modestly affected by a 
second parameter, with there being more evidence for a correlation between
colour residuals and $(B-V)_p$ than with $B_T$, although this conclusion is
greatly dependent on the data for three clusters with [Fe/H] $\approx -1.4$.
As such, the Washington colours may behave with respect to HB morphology in a 
manner that is more similar to that of the integrated $B-V$ colour 
than $U-B$. Nonetheless, perhaps the most valid conclusion is that 
any influence of the second parameter on integrated Washington colours is near 
the limit of the uncertainties in the available measurements.

The implication for the study of extragalactic globular clusters is that in the
Milky Way GC system, none of the commonly used colours $B-V$, $V-I$, $B-I$, or 
$C-T_1$, show deviations from colour-metallicity relations that can be 
systematically related to horizontal branch morphology. Most of the residuals 
that we find in these colours may be more attributable to errors in 
observational measurement or interstellar reddening. A similar conclusion may 
be applicable to the Sloan colour $g-z$ (which has seen increasing use in the 
study of extragalactic globular clusters), since the bandpasses are similar to 
$B-I$. There may be weak horizontal branch effects in $B-V$ and $C-T_1$, but 
these are at the limit of the observational uncertainties. This conclusion
is not without caveats, however, due to (i) the restricted metallicity 
range of Milky Way globular clusters compared to some GC systems in 
external galaxies, and (ii) the ever-lurking second-parameter effect, which 
may behave differently in external systems.

\acknowledgements
We are grateful to Daniel Harbeck for valuable discussions during the 
formative stages of this work.


\begin{thebibliography}{}
\bibitem{} Aaronson, M., Cohen, J. G., Mould, J., Malkan, M.: 1978, 
  ApJ~223, 824
\bibitem{} Armandroff, T. E., \& Da Costa, G. S.; 1991, AJ~101, 1329
\bibitem{} Barmby, P., Huchra, J. P.: 2000, ApJ~531, L29
\bibitem{} Barmby, P., Huchra, J. P., Brodie, J. P., Forbes, D. A., 
 Schroder, L. L., Grillmair, C. J.: 2000, AJ~119, 727 
\bibitem{} Bedin, L. R., Piotto, G., Zoccali, M., Stetson, P. B., Saviane,
 I., Cassisi, S., Bono, G.: 2000, A\&A~363, 159
\bibitem{} Bell, R. A., Dickens, R. J., Gustafsson, B.: 1979, ApJ~229, 604
\bibitem{} Bica, E. L. D., Pastoriza, M. G.: 1983, Ap\&SS~91, 99
\bibitem{} Brocato, E., Castellani, V., Poli, F. M., Raimondo, G.: 2000, 
 A\&A~146, 91
\bibitem{} Brodie, J. P., Huchra, J. P.: 1990, ApJ~362, 503
\bibitem{} Buonanno, R., Caloi, V., Castellani, V., Corsi, C., Fusi Pecci, F.,
 Gratton, R.: 1986, A\&AS~66, 79
\bibitem{} Buonanno, R., Fusi Pecci, F., Cappellaro, E., Ortolani, S., 
 Richtler, T., Geyer, E. H.: 1991, AJ~102, 1005
\bibitem{} Canterna, R.: 1976, AJ~81, 228
\bibitem{} Catelan, M., Borissova, J., Sweigart, A. V., Spassova, N.
 1998, ApJ, 494, 265
\bibitem{} Catelan, M., de Freitas Pacheco, J. A.: 1995, A\&A~297, 345
\bibitem{} Claria, J. J., Minniti, D., Piatti, A. E., Lapasset, E.: 1994a, 
 MNRAS~268, 733
\bibitem{} Claria, J. J., Piatti, A. E., Lapasset, E.: 1994b, PASP~106, 436
\bibitem{} Cohen, J. G., Matthews, K.: 1994, AJ~108, 128
\bibitem{} Covino, S., Pasinetti, L. E., Malagnini, M. L., Buzzoni, A.: 1994,
  A\&A~289, 775
\bibitem{} De Angeli, F., Piotto, G., Cassisi, S., Busso, G., Recio-Blanco, A.,
 Salaris, M., Aparicio, A., Rosenberg, A.: 2005, AJ~130, 116 
\bibitem{} de Freitas Pacheco, J. A., Barbuy, B.: 1995, A\&A~302, 718
\bibitem{} Djorgovski, S., Piotto, G.: 1992, AJ~104, 2112
\bibitem{} Djorgovski, S., Piotto, G., Phinney, E. S., Chernoff, D. F.: 1991,
 ApJ~372, L41
\bibitem{} Forte, J. C., Geisler, D., Ostrov, P. G., Piatti, A. E., Gieren, W.:
 2001, AJ~121, 1992
\bibitem{} Frogel, J. A., Persson, S. E., Cohen, J. G.: 1980, ApJ~240, 785
\bibitem{} Fusi Pecci, F., Ferraro, F. R., Ballazini, M., Djorgovski, S., 
 Piotto, G., Buonanno, R.: 1993, AJ~105, 1145
\bibitem{} Geisler, D.: 1986, PASP~98, 762
\bibitem{} Geisler, D., Claria, J. J., Minniti, D.: 1991, AJ~102, 1836
\bibitem{} Geisler, D., Forte, J. C.: 1990, ApJ~350, L5
\bibitem{} Grundahl, F., Briley, M., Nissen, P. E., Feltzing, S.: 2002, 
 A\&A~385, L14
\bibitem{} Hamuy, M.: 1984, A\&AS~57, 91
\bibitem{} Hanes, D. A., Brodie, J. P.: 1985, MNRAS~214, 491
\bibitem{} Harris, G. L. H., Geisler, D., Harris, H. C., Hesser, J. E.: 1992,
  AJ~104, 613
\bibitem{} Harris, G. L. H., Harris, W. E., Geisler, D.: 2004, AJ~128, 723
\bibitem{} Harris, H. C., Canterna, R.: 1977, AJ~82, 798
\bibitem{} Harris, W. E.: 1974, ApJ~192, L161
\bibitem{} Harris, W. E.: 1996, AJ~112, 1487
\bibitem{} Harris, W. E., Racine, R.: 1979, ARA\&A~17, 241
\bibitem{} Howell, J. H., Guhathakurta, P., Tan, A.: 2000, AJ~119, 1259
\bibitem{} Janes, K. A.: 1975, ApJS~29, 161
\bibitem{} Johnson, J. A., Bolte, M.: 1998, AJ~115, 693
\bibitem{} Kurth, O. M., Fritze-von Alvensleben, U., Fricke, K. J.: 1999, 
 A\&AS~138, 19
\bibitem{} Lee, H.-C., Yoon, S.-J, Lee, Y.-W.: 2000, AJ~120, 998
\bibitem{} Lee, M. G., Geisler, D.: 1993, AJ~106, 493
\bibitem{} Leitherer, C., et al.: 1996, PASP~108, 996
\bibitem{} Maraston, C.: 1998, MNRAS~300, 872
\bibitem{} Maraston, C., Thomas, D.: 2000, ApJ~541, 126
\bibitem{} Maraston, C.: 2005, MNRAS~362, 799
\bibitem{} McClure, R. D.: 1973, in: C. Fehrenbach, B.~E. Westerlund (eds.),
 {\it Spectral Classification and Multicolor Photometry, IAU Symposium No.~50},
  Riedel, Dordrecht, p. 162
\bibitem{} McLaughlin, D. E., Secker, J., Harris, W. E., Geisler, D.: 1995, 
 AJ~109, 1033
\bibitem{} Moehler, S., Sweigart, A. V., Catelan, M.: 1999, A\&A~351, 519
\bibitem{} Momany, Y., Piotto, G., Recio-Blanco, A., Bedin, L. R., Cassis, S.,
 Bono, G.: 2002, ApJ~576, L65
\bibitem{} Newell, B., Sadler, E. M.: 1978, ApJ~221, 825
\bibitem{} Osborn, W.: 1973, ApJ~186, 725
\bibitem{} Ostrov, P. G., Forte, J. C., Geisler, D.: 1998, AJ~116, 2854
\bibitem{} Peterson, R. C., Carney, B. W., Dorman, B., Green, E. M., Landsman, 
 W., Liebert, J., O'Connell, R. W., Rood, R. T.: 2003, ApJ~588, 299
\bibitem{} Philip, A. G. D., Cullen, M. F., White, R. E.: 1976, {\it UBV 
 Color-Magnitude Diagrams of Galatic Globular Clusters, Dudley Observatory 
 Report No. 11}, Dudley Observatory Press, Albany 
\bibitem{} Press, W. H., Teukovsky, S. A., Vetterling, W. T., Flannery, B. P.:
 1999, {\it Numerical Recipes in Fortran 77, 2nd ed.}, Cambridge University 
 Press, Cambridge
\bibitem{} Puzia, T. H., et al.: 2005, A\&A~439, 997
\bibitem{} Racine, R.: 1973, AJ~78, 180
\bibitem{} Reed, B. C.: 1985, PASP~97, 120
\bibitem{} Reed, B. C., Hesser, J. E., Shawl, S. J.: 1988, PASP~100, 545
\bibitem{} Rey, S.-C., Yoon, S.-J., Lee, Y.-W., Chaboyer, B., Sarajedini, A.: 
 2001, AJ~122, 3219
\bibitem{} Rich, R. M., Sosin, C., Djorgovski, S. G., Piotto, G., King, I. R.,
 Renzini, A., Phinney, E. S., Dorman, B., Liebert, J., Meylan, G.: 1997,
 ApJ~484, L25  
\bibitem{} Rosenberg, A., Saviane, I., Piotto, G., Aparicio, A.: 1999, AJ~118, 
 2306
\bibitem{} Rutledge, G. A., Hesser, J. E., Stetson, P. B., Mateo, M.,
 Simard, L., Bolte, M., Friel, E. D., Copin, Y.: 1997a, PASP~109, 883
\bibitem{} Rutledge, G. A., Hesser, J. E., Stetson, P. B.: 1997b, PASP~109, 
 907
\bibitem{} Sandage, A., Wildey, R.: 1967, ApJ~150, 469
\bibitem{} Schiavon, R. P., Rose, J. A., Courteau, S., MacArthur, L. A.: 
 2004, ApJ~608, L33
\bibitem{} Sil'chenko, O. K.: 1983, SvAL~9, 145
\bibitem{} Spearman, C.: 1904, Am. J. Psychol.~15, 72
\bibitem{} Sung, H., Lee, S.-W.: 1987, JKAS~20, 63
\bibitem{} Tripicco, M. J., Bell, R. A.: 1991, AJ~102, 744
\bibitem{} van den Bergh, S.: 1967, AJ~72, 70
\bibitem{} Walker, A. R.: 1999, AJ~118, 432 
\bibitem{} White, R. E., Shawl, S. J.: 1987, ApJ~317, 246 
\bibitem{} Worthy, G.: 1994, ApJS~95, 107
\bibitem{} Yi, S.: 2003, ApJ~528, 202
\bibitem{} Zinn, R., West, M.: 1984, ApJS~55, 45
\end{thebibliography}
\end{document}